\documentclass[11pt,preprint]{aastex}
\usepackage{epsfig}

\newcommand{\um}{\ensuremath{\mu}m}
\newcommand{\degree}{\ensuremath{\!\symbol{23}\!\!}}
\newcommand{\deltat}{\ensuremath{\Delta t}}
\newcommand{\radiusvec}{\mathbf{r}}
\newcommand{\unitradiusvec}{\mathbf{\hat{r}}}
\newcommand{\velocityvec}{\mathbf{v}}

\newcommand{\etal}{et al.}
\newcommand{\sma}{semi-major axis}
\newcommand{\smas}{semi-major axes}
\newcommand{\erid}{$\epsilon$~Eridani}
\newcommand{\pr}{Poynting-Robertson}

\shorttitle{Numerical Modelling of Dusty Debris Disks}
\shortauthors{Deller \& Maddison}

\slugcomment{Accepted for publication in the Astrophysics Journal}

\begin{document}

\title{Numerical Modelling of Dusty Debris Disks}
\author{A. T. Deller and S. T. Maddison}
\affil{Centre for Astrophysics and Supercomputing, Swinburne University, PO Box 218, Hawthorn, VIC 3122, AUSTRALIA }

\begin{abstract}
  Infrared and submillimetre observations of nearby Vega-like
  stars have revealed a number of clumpy, asymmetric dust debris
  disks. Previous studies using semi-analytical and numerical methods
  have suggested planetary companions of various mass as the likely
  cause of most examples of disk asymmetry. In this paper, we modify an
  N-body symplectic gravitational integrator to include radiation
  terms and conduct medium-resolution parameter searches to identify
  likely planetary candidates in observed Vega-like systems.  We also
  present high resolution models of Vega and \erid,
  comparing our results to those of previous authors, and a new model
  for Fomalhaut.
\end{abstract}

\keywords{circumstellar matter -- methods: N-body simulations -- planetary
systems -- stars: individual (Vega, \erid, Fomalhaut)}

%-------------------------------------------------------
\section{Introduction}

Dust disks composed of particles in the micron to centimetre range are
known to exist around many stars that exhibit an infrared excess, such
as Vega, $\beta$~Pictoris, \erid \, and others \citep[see e.g.][]{zuc01}.
Particles of this size are affected significantly by
solar radiation and corpuscular (solar wind) forces, with
Poynting-Robertson (PR) and solar wind drag resulting from the moving
particle's absorption and reemission of solar energy.  Over time,
these dissipative forces caused by radiation and solar wind act to
reduce a dust particle's \sma\ and eccentricity, and the
particle eventually spirals into the central star \citep{burns79}.
Since the age of most of these
systems exceeds the expected lifetime of a small dust particle, the
particles must be continually replenished.  The suspected mechanism
is a collisional cascade involving primordial planetesimal and
smaller sized bodies \citep{backman93,wyatt02} as is believed to occur in the
Edgeworth-Kuiper Belt \citep[e.g.][]{backman95,land02}.  The destructive process
which leads to dust formation has led these disks to be known as `debris
disks'.

Of the debris disks near enough to Earth to be spatially resolved,
most show a significant degree of asymmetry (e.g. Vega: \citealt{holland98};
\erid: \citealt{greaves98}).  The commonly held view
is that gravitational interactions with an unseen embedded planet are
responsible for most cases of disk structure \citep{oz00,qt02,whk02}.  Alternative possible
explanations have been considered for specific systems, such as
disturbance by a passing star \citep[e.g. HD141569:][]{clampin03},
a recent large cometary/planetesimal collision \citep[e.g. Fomalhaut:][]{wyatt02}
and contamination by a background feature such as a
distant galaxy \citep[e.g. Vega:][]{whk02}.

A planetary companion has several effects on an initially uniform
debris disk.  Particles near the planet are often ejected due to a
close encounter -- an effect which is enhanced as the planet's mass
increases.  Since particles interior to the planet spiral increasingly
rapidly into the central star, an interior cleared zone is usually 
present \citep{lz99}.  However, some particles spiralling in
towards the star may become trapped in a series of Mean Motion
Resonances (MMRs) with the planet.  An MMR exists when the ratio of
orbital periods for two objects orbiting a central mass is equal to
$m:n$, where $m$ and $n$ are integers.  The order of an MMR is given
by the quantity $\left| m-n \right|$.  Earth has a dust ring formed by dust
particles in MMRs \citep{dermott94}, and Neptune is suspected to do
the same to Kuiper Belt dust \citep{lz99}.

Whilst locked into an MMR with a planet, the angular momentum of the
particle lost to \pr\ and solar wind drag is balanced by the resonant
forcing of the planet's gravitation.  A particle trapped in an MMR can
extend the particle's lifetime to many times the
value predicted by \pr\ and solar wind drag alone.  The resultant
`pile-up' of particles at discrete \smas\ tends to cause prominent
radial and angular structure in the debris disk \citep{kh03}.

In multiple planet systems, it is the outermost planet that generally
dominates the structure of the resulting debris disk.  Neptune is
suspected to play this role in interactions with Kuiper Belt dust
\citep{lz99}. The interior planets, however, can have significant
effects on particles which `leak past' the outermost planet. In the solar system, for example, Jupiter and
Saturn eject many dust particles which drift interior to Neptune \citep{lz99}.
For simplicity, we investigate only single
planet systems and assume that the effects of any interior planets
which may be present is negligible outside the orbit of the outermost
planet.

In this paper, we numerically simulate a large number of debris disk systems, creating a synthetic debris disk catalogue.  Our numerical method is described in Section~\ref{sec:numericmeth}, and the
testing of the code is detailed in Section~\ref{sec:verifycode}.  We present a small sample of the results from the synthetic catalogue in Section~\ref{sec:synthcat}, and use the catalogue to select planetary configurations which could be responsible for the observed systems of Vega, \erid\ and Fomalhaut.  The selected systems are then simulated in higher detail in Section~\ref{sec:compareprev}, and the results compared to observational data.  Our conclusions are presented in Section~\ref{sec:conclusions}.

%-------------------------------------------------------

\section{Numerical Method}
\label{sec:numericmeth}

We model the evolution of debris disks using an N-body integrator that includes the effects of radiation pressure, \pr\ and solar wind drag.  In this section we discuss the modifications made to a symplectic integrator to account for the additional forces, as well as our numerical techniques for creating debris disks and recording particle positions.

\subsection{The RMVS3 Integrator}
Since the general problem of a particle moving under the influence of
gravitational and radiation forces is nonlinear, it is necessary
to numerically integrate the orbits of all particles.  Commonly used
integrators include Runge-Kutter, Burlisch-Stoer and Mixed Variable
Symplectic (MVS) algorithms.  The MVS integrator developed by
\citet{wisdom91} is normally the fastest, but has the
disadvantage of being unable to follow close encounters between
particles and planets, which are crucial to debris disk evolution.
This drawback has been overcome with the use of integrators such as
RMVS3 \citep{ld94} and SyMBA \citep{dll98}, which
are based on the \citet{wisdom91} scheme but can integrate close encounters.  
In this work we have
modified RMVS3 to include radiation and solar wind forces which affect
debris disk particles.  Such an approach was used by \citet{mm02},
who modify a variant of SyMBA. (Note that SyMBA and RMVS3
handle close encounters between particles and planets in different ways.)

RMVS3 assumes that test particles are both massless and
collisionless -- the problem of dust collisions is discussed in
Section~\ref{subsec:collisionalproc}.  The code requires units such that
the gravitational constant $G = 1$ and so we have chosen distance,
time and mass units to be 1~AU, 1~year and $M_{\odot}/(4\pi^{2})$ respectively.
Throughout this paper we shall use $a$ for the semi-major axis,
$e$ for eccentricity, and $i$ for inclination, with the subscripts
${tp}$, ${pb}$ and ${pl}$ referring to a test particle, parent body and planet respectively.

As developed in \citet{wisdom91} and described in \citet{ld94}, RMVS3 expands the
Hamiltonian of a test particle into two integrable components given by
\begin{equation}
H = H_{kep} + H_{dist} \, ,
\end{equation}
where $H_{kep}$ represents the Keplerian motion around the central
star, and $H_{dist}$ represents the perturbances on a test particle
resulting from the planet(s).  Using a timestep \deltat, the second
order approximation implemented in RMVS3 consists of applying the
disturbance Hamiltonian for \deltat/2, applying the Keplerian
Hamiltonian for \deltat, and then applying the disturbance Hamiltonian
again for \deltat/2.  The approximation is accurate as long as the
planetary disturbances are small compared to the Keplerian evolution.
This assumption of small disturbances does not hold when a particle
suffers a close encounter with a planet.  In this situation, since the
gravitational effect of the planet on the particle exceeds that of the star,
the roles of planet and star are reversed and the Keplerian portion of the
Hamiltonian is taken to occur around the planet and the star is
relegated to the role of disturbance \citet{ld94}.  Although a smaller timestep
is used during a close encounter, the duration of an encounter is typically small enough 
that the simulation proceeds rapidly. In the intermediate region where planetary
 and stellar forces are comparable, the timestep is also reduced but the integration remains
 heliocentric.

\subsection{Addition of Radiation and Solar Wind}
The magnitude of the radiation force on a particle is given by
\begin{equation}
\left| F_{rad} \right| = \frac{L A Q_{PR}}{4\pi cr^{2}} \, ,
\end{equation}
where $L$ is stellar
luminosity, $A$ is the particle cross-sectional area, $Q_{PR}$ is the
radiation pressure coefficient, $c$ is the speed of light and $r$ is
the heliocentric distance.  It is standard practice to describe the
strength of the radiation force on a given particle as a fraction of
the gravitational force on a particle, such that
\begin{equation}
F_{rad} = \beta F_{grav} \, .
\end{equation}
The constant $\beta$ is depends on the particle size and composition, as
well as the stellar mass and luminosity, but is independent of the
particle's heliocentric distance. The magnitude of \pr\ drag is
proportional to $\beta$. The ratio of solar wind drag to \pr\ drag is
dependent on particle size, but is relatively constant over the range
of particles usually considered \citep[$s > 0.5 \mu$m, ][]{burns79}, so the combined drag effects of PR
and solar wind can be expressed using a single parameter given by
\begin{equation}
\beta_{sw} = \beta \left( 1 + sw \right) \, ,
\end{equation}
where $sw$ is the ratio of solar wind drag to \pr\ drag \citep{burns79}.
Following \citet{mm02}, we use a constant value of
$sw=0.35$ in all our simulations.

The acceleration on a particle due to solar wind drag and radiation,
ignoring terms of order $(v/c)^{2}$\ and higher, is given by
\begin{equation}
\frac{d^{2} \radiusvec}{dt^{2}} = F_{grav} \left[ \beta \unitradiusvec -
  \frac{\beta_{sw}}{c}\left( v_{r} \unitradiusvec + \velocityvec \right)
  \right] \, .
\end{equation}
The first term in this expression is the result of the
outwardly-directed radiation pressure, which causes the net
acceleration in the radial direction to be reduced to
$(1-\beta) F_{grav}$.  The remaining two terms (which are smaller by a
factor of $v/c$) represent the \pr\ and solar wind drag.  \pr\ drag results
from the particle's motion relative to the radiation source, which
causes a component of the radiation force to oppose the particle's
motion. During normal particle movement, the first term is accounted for
in the Keplerian Hamiltonian by reducing the central object mass to the
`apparent' value, while the other terms are added to the disturbance
Hamiltonian.  During a close encounter with a planet, the entire expression
is added to the disturbance Hamiltonian.

\subsection{Particle Initialisation}

All simulations commence with a central star, an orbiting planet and a number of test particles distributed around the star with a fixed value of $\beta$.  The star is fixed at the origin and has mass represented by $M_{\star}$.  The orbiting planet is described by $M_{pl}$, $a_{pl}$ and $e_{pl}$, representing the planet's mass, \sma\ and eccentricity respectively.

Test particles are assumed to be released from larger `parent bodies' (which are unaffected by radiation pressure) at the beginning of the simulation.  This is analogous to the creation of dust through the collision of larger rocky bodies in a debris disk.  The concept of a parent body is used only to set initial orbital parameters for test particles; they play no part in the simulation after initialisation.
For each simulation, the initial distribution of parent bodies is specified by a range of semi-major axis, eccentricity and inclination values, given by $a_{pb}$, $e_{pb}$ and $i_{pb}$ respectively.  The arguments of perihelion, longitudes of ascending nodes and mean anomalies for the parent bodies are allocated randomly.  A test particle is then created from each parent body, with semi-major axis, eccentricity and inclination referred to by $a_{tp}, e_{tp}$ and $i_{tp}$ respectively.  Since the test particles are affected by radiation pressure while the parent bodies are not, the test particles will have different orbital parameters from their parent bodies.  In particular, the test particles will have increased \smas\ and changed eccentricities, given by

\begin{equation}
a_{tp} = a_{pb}\frac{1-\beta}{1-2a_{pb}\beta/r} ,
\label{eqn:axischange}
\end{equation}
\begin{equation}
e_{tp} = \left( 1 - \frac{(1-2a_{pb}\beta/r)(1-e_{pb}^{2})}{\left( 1-\beta \right)^{2}} \right)^{1/2} .
\label{eqn:eccchange}
\end{equation}

\subsection{Particle Recording}
\label{subsec:recording}
Simulating a sufficient number of test particles to resolve the disk structure
at all times is computationally expensive and an alternative method, used by
\citet{lz99} and \citet{mm02}, is to simulate a
smaller number of particles and regularly record their positions, which can then be summed to produce distribution maps.  A simulation is terminated after a fixed time or when all particles have been destroyed through either ejection via close encounters with planets or accretion onto the central star.  In this way, long lived particles (trapped in resonances) contribute more to the final dust distribution.  If we assume the dust distribution is ergodic and that
the total dust mass is constant (i.e. the dust production rate equals the
loss rate), this approach should give an accurate representation of the
actual `steady-state' disk structure.

In previous studies \citep[e.g.][]{mm02} the particle locations have been transformed into a
reference frame which rotates with the outermost planet, but this approach
is not appropriate for planets with eccentric orbits.  For this reason, we
instead ensure that the particle recording takes place with the planet in
specific orbital phases.  In this way a number of `snapshots' of the system
are taken with the planet in varying locations, allowing the
effect of planetary phase to be studied \citep{whk02,kh03}.  For example, many simulations utilised a planet 
with orbital period 300 years and a particle recording time of 1000 years, resulting in 
particle distributions for three different planetary phases.

It should be noted that there are some limitations to this procedure of
simulating a debris disk using a small number of test particles.
As discussed by \citet{mm02}, because both the probability
of being trapped in a particular resonance and the time spent in a resonance
 depend sensitively on initial conditions, using a limited number of particles
may over-exaggerate the importance of a particular resonance.  We have tried
to overcome this problem by using an order of magnitude more particles than previous simulations
\citep[e.g.][]{qt02,mm02} in high resolution simulations (see following sections).  An additional problem with
this procedure is that all of the test particles are released from their parent body at the same time, with a specific planetary phase.  In reality, dust particles would be released continuously, at random planetary phases. This effect, however, is not likely to have a significant
impact, since the test particles are typically released a long way
from the planet and are distributed randomly around the star.

Setting a fixed maximum simulation time may also be problematic.  For the small particle number approach to be accurate, the integration should be continued until all test particles have been destroyed.  The termination of a simulation at a predetermined time means that some particles may still be active at the end of the simulation.  One could argue, however, that extremely long-lived particles are unlikely to exist due to destruction via particle collisions.  Since RMVS3 is a collisionless code, the destruction of particles in this way is not accounted for.  We investigate these effects in more detail in Section~\ref{subsec:collisionalproc}.

%-------------------------------------------------------
\section{Test Calculations}
\label{sec:verifycode}
There is no general solution for the motion of a particle under the
influence of radiation and gravitational forces in a system
comprising a star and one or more planets.  However, analytic
solutions and analytic approximations exist for particle motion in the two-body  \citep{wyatt50}
and the circularly restricted three-body \citep{lz97} problems respectively.
These specific cases can be used to check the accuracy of the
modified RMVS3 code.

To ensure that the addition of radiation and
solar wind forces did not effect the normal running of the code, we
ran a series of tests in which $\beta$ was set to zero, and compared 
the results to identical tests run using the original RMVS3.  The
results were indistinguishable in all cases.

\subsection{The 2-body problem}
The time rate of change of a test particle's orbital elements in the
radiation-modified two body problem are given in \citet{mm02} (following the work of 
\citet{wyatt50,burns79}), and are equal to
\begin{eqnarray}
\left( \frac{da}{dt} \right) & = & -\frac{(1+sw)\beta M_{\star}}{c}
\frac{\left( 2+3e^{2} \right)}{a \left( 1 - e^{2} \right) ^{3/2}} \, ,\\
\left( \frac{de}{dt} \right) & = & -\frac{5(1+sw)\beta M_{\star}}{2c}
\frac{e}{a^{2} \left( 1 - e^{2} \right) ^{1/2}}             \, ,    \\
\left( \frac{di}{dt} \right) & = & 0 \, .
\end{eqnarray}
where $M_{\star}$ is the mass of the central star, and $c$ is the speed of light.

\begin{figure}
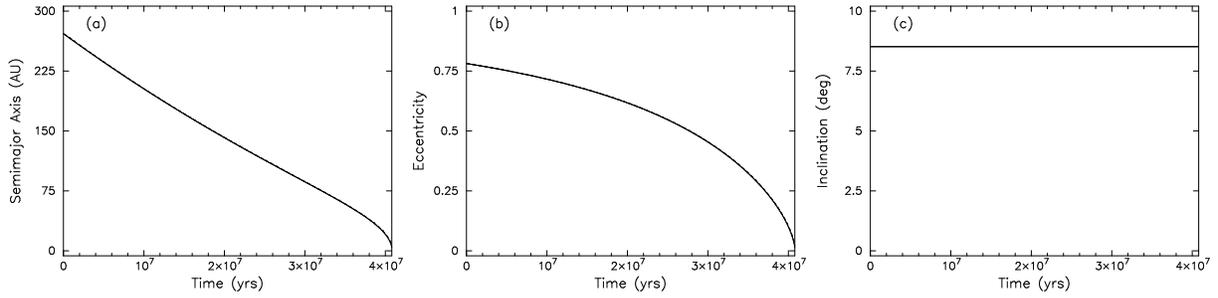

\begin{center}
\begin{tabular}{c}
\psfig{file=f1a.eps, width=0.23\textwidth, angle=270}
\psfig{file=f1b.eps, width=0.23\textwidth, angle=270}
\psfig{file=f1c.eps, width=0.23\textwidth, angle=270}
\end{tabular}
\caption{Time evolution of a single test particle orbiting a solar mass star
  subject to \pr\ and solar wind drag. Plotted are the evolution of
  (a) \protect$a_{tp}$,
  (b) \protect$e_{tp}$, and
  (c) \protect$i_{tp}$
  for a particle with \protect$\beta=0.15$ and $sw$ = 0.35.  The analytic
  solution (dashed line) and actual particle path (solid line) are coincident.}
\label{fig:twobody}
\end{center}
\end{figure}

To test the code against the two--body solutions, we ran a simulation with a single $\beta=0.15$ 
test particle placed in orbit around
a solar-mass star, released from a parent body with $a_{pb} = 250$\ AU, $e_{pb} = 0.8$ and
$i_{pb} = 7.6\degree$.  The particle's argument of
perihelion, longitude of ascending node and mean anomaly were selected
randomly. The analytical and numerical results plotted in Figure~\ref{fig:twobody} are
in excellent agreement.

\subsection{The restricted 3-body problem}
The addition of a planet greatly complicates the motion of the test
particle, but \citet{lz97} have derived expressions for the time
evolution of a particle's orbital elements whilst in an MMR with a
zero eccentricity planet.  These expressions are a second order approximation,
valid only for low test particle eccentricities and inclinations, and low order
resonances, and are given by
\begin{eqnarray}
e^{2} & = & \left( e_{0}^{2} - ({K-1})/{3}) \right)
             \exp{(3At/K)}+ ({K-1})/{3} \, , \label{eqn:mmr1} \\
i     & = & i_{0} \exp{(-At/4)} \, , \label{eqn:mmr2}
\end{eqnarray}
where $A = (2\beta_{sw}M_{\star})/(a^{2}c)$, $K = m/n$, $e_0$ and $i_0$ represent the particle's initial eccentricity and inclination, and $m$ and $n$ are the integers specifying the resonance.  It should be
noted that the location of resonances is shifted from their normal location
due to the presence of the radiation force, which causes particles at a
given \sma\ to orbit more slowly due to the reduced effective
gravitational force.  The \sma\ of a particle locked in an $m:n$
resonance with a planet is given by
\begin{equation}
a_{tp} = a_{pl}\left( 1-\beta \right)^{1/3}\left( m/n \right)^{2/3} \, .
\end{equation}

\begin{figure}
\plotone{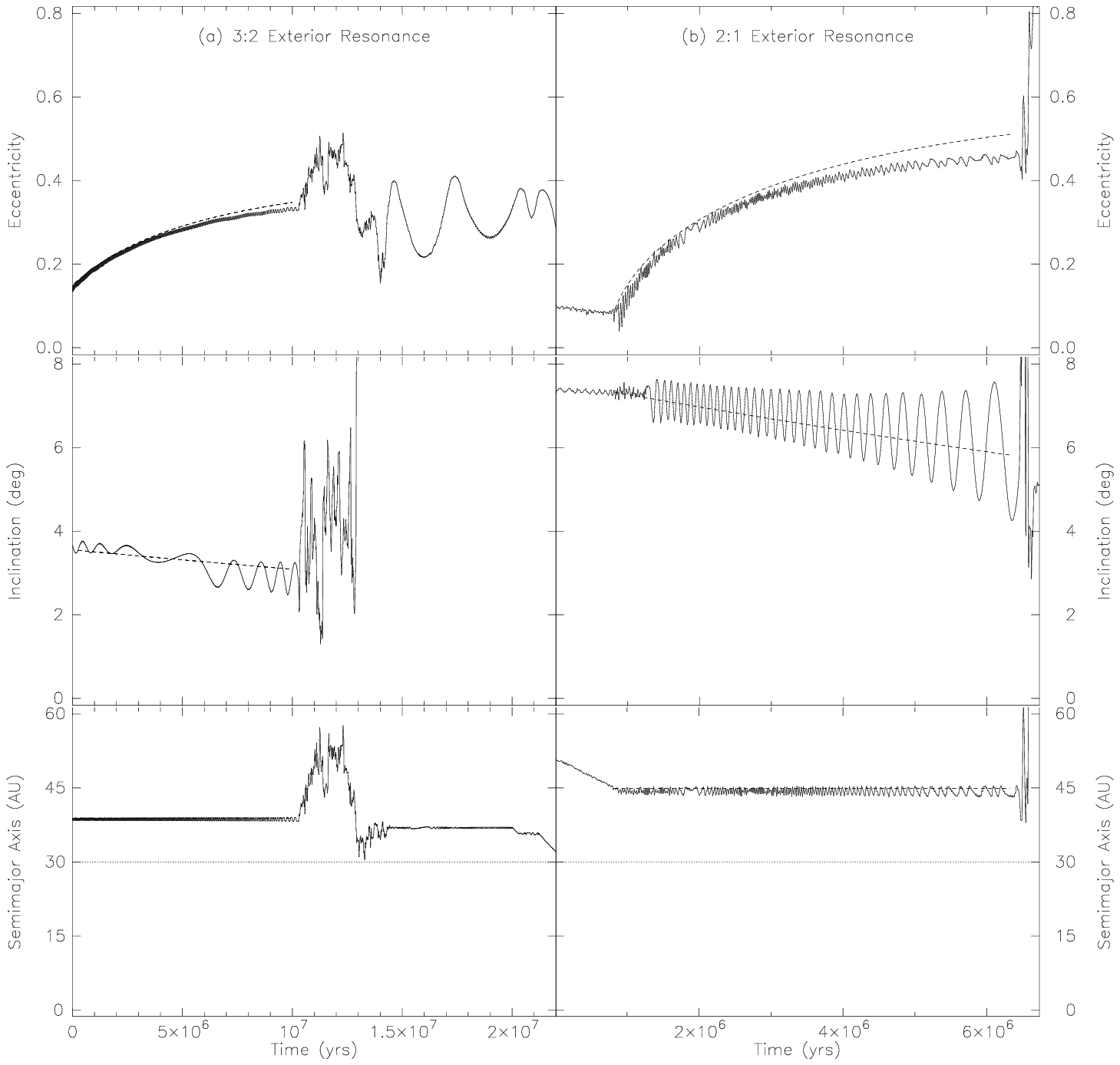}
\caption{Evolution of a single test particle whilst in an MMR with a planet.
The particle evolution is shown as a solid line, and the analytic solution is
shown as a dashed line.  The planetary \sma\ is shown as a dotted line in the
bottom two panels.
(a) Particle with $\beta=0.05$, in a 3:2 resonance with a $0.05 M_{Jup}$ mass planet.  The particle is trapped for about $10^{7}$ years.
(b) Particle with $\beta=0.2$, in a 2:1 resonance with a Jupiter mass planet.  The particle is trapped for about 5 million years.
In both cases $sw=0.35$.}
\label{fig:threebody}
\end{figure}

We start at time $t=0$ with a single particle placed exterior to the resonance
in orbit around a one solar mass star and a planet at 30~AU.  We follow the
orbital evolution of the test particle as it passes through the MMR until it
is ejected from the system, and compare the results with
equations~\ref{eqn:mmr1} and \ref{eqn:mmr2}.
%
%We have applied these analytic solutions to particles evolving in a
%number of different MMRs in different planetary systems.
%
We show the results for two different $\beta$ values.  In
Figure~\ref{fig:threebody}a, a single $\beta = 0.05$ particle is trapped in
a $3:2$ resonance with a Neptune mass planet, while in
Figure~\ref{fig:threebody}b a particle with $\beta=0.2$ is trapped in a
$2:1$ resonance with a Jupiter mass planet. The particles' orbital elements
are well described until their eccentricity and inclination become too
large and the analytic solution becomes invalid. The results shown in
Figure~\ref{fig:threebody} are only a small sample of the large number of
tests carried out to ensure the accuracy of the code.

%-------------------------------------------------------
\section{A Synthetic Debris Disk Catalogue}
\label{sec:synthcat}
Previous studies by \citet{kh03} have predicted the general effects of varying planetary mass and eccentricity on the dust distribution in a debris disk.  We have generated a synthetic catalogue of debris disks which provides numerical agreement to the theoretical predictions and can be used as a guide when interpreting observed systems.  Here, we present a sample of results obtained from modelling over 300 disks, showing the effects of planetary mass and eccentricity, dust composition ($\beta$) and initial dust distribution on the disk structure.  By generating simulated observations of these simulations, we show that even observations at a low resolution can distinguish between different planet and dust combinations.

For the creation of our synthetic catalogue, we allowed five parameters to vary: $M_{pl}, e_{pl}$, $\beta$, $a_{pb}$ and $e_{pb}$.  All other parameters were fixed as follows: $M_{\star} = 1 M_{\odot}$, $0\degree < i_{pb} < 8\degree$, $n_{tp} = 500$, and planetary period = 300 years (yielding $a_{pl} \sim 44.8$~AU). Table~\ref{tab:parameters} summarises the parameter space we have investigated.  Of the 600 possible combinations of our five free parameters, approximately 300 disks were simulated.  Subsets of the parameter space which were fully explored include $\beta = 0.1$, $M_{pl} = 0.05 M_{Jup}$ and $M_{pl} = 1 M_{Jup}$.

\clearpage
\begin{center}
\begin{table}
\begin{center}
\begin{tabular}{cccll} \hline
  &   &  & \ \ \ \ \ $a_{pb}$/$a_{pl}$   & \ \ \ \ \ \ \ \ $e_{pb}$ \\
$\mathbf M_{pl}/M_{Jup}$ & $\mathbf e_{pl}$ & $\mathbf \beta$
  & \ \ $\mathbf [a_{min}, a_{max}]$
  & \ \ $\mathbf [e_{min}, e_{max}]$ \\ \hline
0.01 &   0.0  &  0.01  & [1.0, 1.4] (close) & [0.0, 0.2] (low)    \\
0.05 &   0.1  &  0.1   & [1.4, 2.0] (mid)   & [0.1, 0.5] (medium) \\
0.2  &   0.3  &  0.2   & [2.0, 3.5] (far)   &                     \\
1.0  &   0.5  &  0.4   &                    &                     \\
3.0  &   0.7  &        &                    &                     \\ \hline
\end{tabular}
\caption{The range of parameter values used in generating our synthetic catalogue.  Note that while this parameter space was not completely explored, approximately 300 of the 600 possible disks were simulated.  Fully explored parameters (where the chosen parameter was simulated with all possible combinations of the other parameters) include $\beta=0.1$, $M_{pl}=0.05 M_{Jup}$ and $M_{pl}=1 M_{Jup}$.}
\label{tab:parameters}
\end{center}
\end{table}
\end{center}
\clearpage

Each simulation was run for $6\times10^{7}$ years, corresponding
to $2\times10^{5}$ planetary orbits.  Particle positions were recorded every
1000 years, allowing observation of the system in 3 different planetary
phases.  The maximum and minimum allowable heliocentric radii of the test
particles were 700 and 2 AU respectively.

At the end of each simulation, the recorded particle positions (in the three
orbital phases) were binned in the $XY$-plane.  The particle distribution in the $XY$-plane was then plotted
for each particular planetary phase.  We also
binned particle \smas\ and plotted the \sma\ occupancy versus semi-major axis, which shows the location and
strength of MMRs with the planet.

In order to produce synthetic observations of the systems, the particle
distribution plots were weighted by estimated total disk mass and the dust
emissivity at 850~$\mu$m, assuming the dust to be a perfect blackbody.  This
emissivity plot was then convolved with a two dimensional Gaussian with a
full width half maximum (FWHM) of 40~AU, to simulate observations of the
system with limited resolution.  The synthetic observations can also be plotted from various disk viewing
angles.  This procedure assumes that the disk is optically thin, which is a reasonable assumption
for disks containing little to no gas, as is the case for most Vega-like
systems \citep{liseau99}.

\begin{figure}
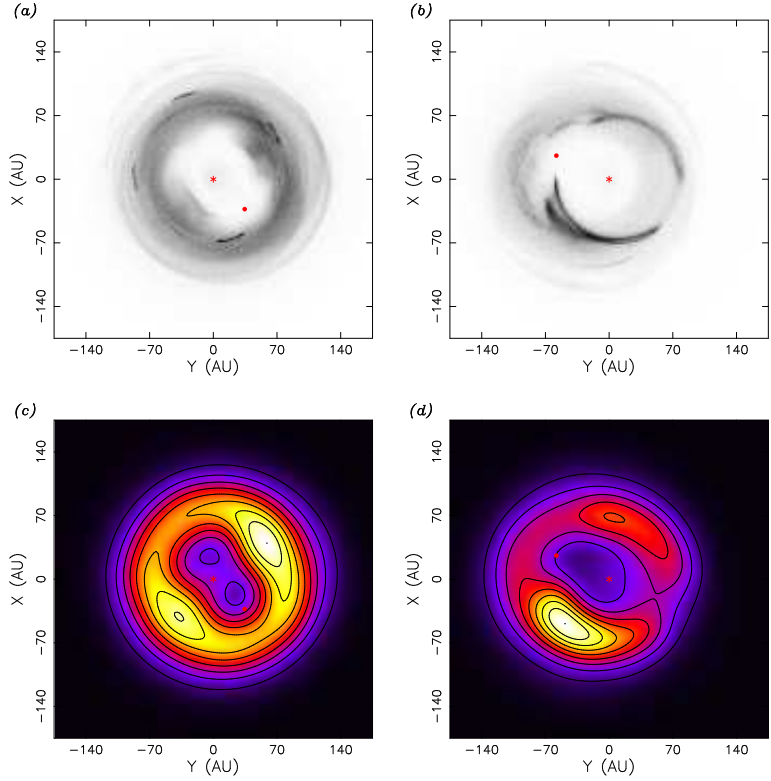

\begin{center}
\begin{tabular}{cc}
\psfig{file=f3a_col.eps, width=0.3\textwidth, angle=270} &
\psfig{file=f3b_col.eps, width=0.3\textwidth, angle=270} \\
\psfig{file=f3c_col.eps, width=0.3\textwidth, angle=270} &
\psfig{file=f3d_col.eps, width=0.3\textwidth, angle=270}
\end{tabular}
\caption{Simulations of a system containing a single Jupiter mass planet and 500 test particles after $6 \times 10^{7}$ years, or 20,000 orbital periods.  In each case, the planet is located at $a_{pl} \sim 44.8$~AU, test particles were released from parent bodies with $1.4<a_{pb}/a_{pl}<2.0$, $0<e_{pb}<0.2$ and $0<i_{pb}<8$\degree and $\beta = 0.1$.  The star and planet are represented by a filled star and circle respectively (coloured red in the online edition of the Journal).
Plots (a) and (b) show the dust distributions generated when $e_{pl} = 0.1$ and 0.5 respectively.
Plots (c) and (d) show simulated observations of the disks (with beam FWHM 40 AU) for $e_{pl} = 0.1$ and 0.5 respectively.  See the electronic edition of the Journal for a colour version of this figure.
}
\label{fig:synthcatjup}
\end{center}
\end{figure}

\begin{figure}
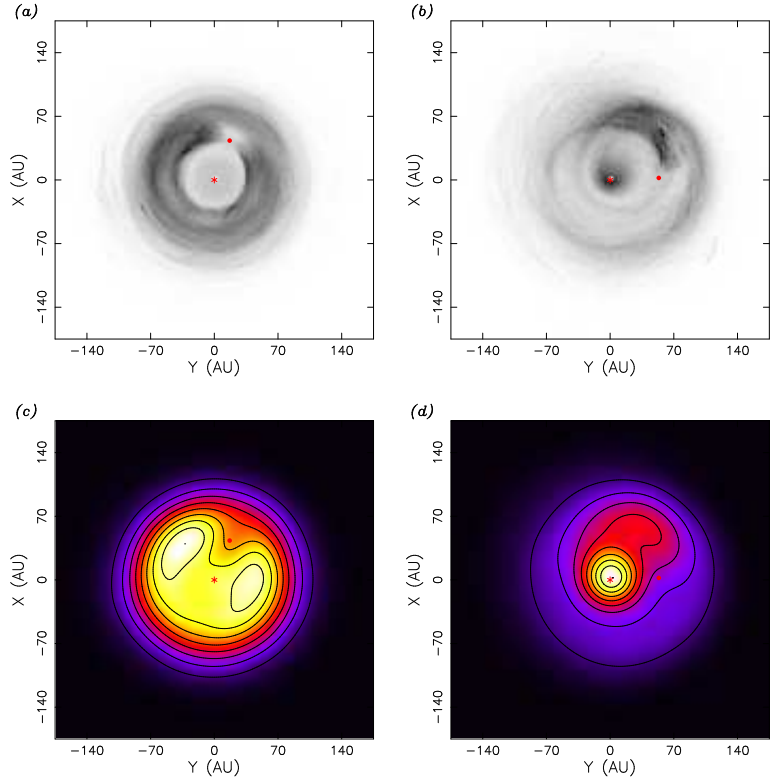

\begin{center}
\begin{tabular}{cc}
\psfig{file=f4a_col.eps, width=0.3\textwidth, angle=270} &
\psfig{file=f4b_col.eps, width=0.3\textwidth, angle=270} \\
\psfig{file=f4c_col.eps, width=0.3\textwidth, angle=270} &
\psfig{file=f4d_col.eps, width=0.3\textwidth, angle=270}
\end{tabular}
\caption{Simulations of a system containing a single Neptune mass planet and 500 test particles after $6 \times 10^{7}$ years, or 20,000 orbital periods.  In each case, the planet is located at $a_{pl} \sim 44.8$~AU, test particles were released from parent bodies with $1.4<a_{pb}/a_{pl}<2.0$, $0<e_{pb}<0.2$ and $0<i_{pb}<8$\degree and $\beta = 0.1$.
Plots (a) and (b) show the dust distributions generated when $e_{pl} = 0.1$ and 0.5 respectively.
Plots (c) and (d) show simulated observations of the disks (with beam FWHM 40 AU) for $e_{pl} = 0.1$ and 0.5 respectively.  See the electronic edition of the Journal for a colour version of this figure.
}
\label{fig:synthcatnep}
\end{center}
\end{figure}

The complete synthetic catalogue is available for viewing online\footnote{\tt http://astronomy.swin.edu.au/debrisdisks/}.
Here we present a sample of results from our catalogue for Jupiter and Neptune mass planets with $e_{pl} = 0.1$ and $0.5$ in Figures~\ref{fig:synthcatjup} and \ref{fig:synthcatnep}.  These simulations use $\beta=0.1$, and test particles released from parent bodies with orbital elements
$0<e_{pb}<0.2$, $1.4<a_{pb}/a_{pl}<2.0$ and $0<i_{pb}<8\degree$.

As can be seen in Figures~\ref{fig:synthcatjup} and \ref{fig:synthcatnep},
the mass and eccentricity of a planet dramatically alters the resulting dust
distribution.  The Jupiter-mass planet clears a central cavity, which is not
as pronounced in the Neptune-mass case. In the low planetary eccentricity case, a relatively smooth dust ring is present external to the planet.  When the planetary eccentricity is increased, the smooth ring vanishes, replaced by prominent arcs or clumps of dust emission.  Similar trends were observed in the synthetic catalogue with planets of different masses.  These results closely resemble the predictions made by \citet{kh03} for the four cases of: (a) low mass, low eccentricity planets, (b) low mass, moderate eccentricity planets, (c) high mass, low eccentricity planets, and (d) high mass, moderate eccentricity planets.

\begin{figure}
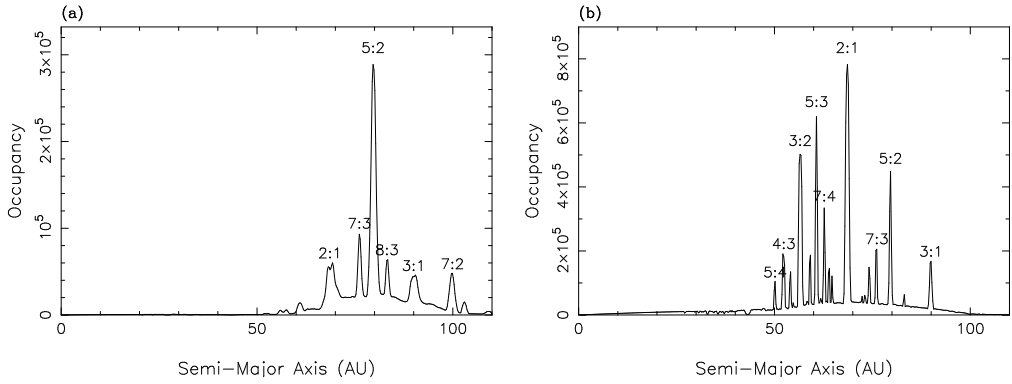

\begin{center}
\begin{tabular}{cc}
\psfig{file=f5a.eps, width=0.3\textwidth, angle=270} &
\psfig{file=f5b.eps, width=0.3\textwidth, angle=270}
\end{tabular}
\caption{
Resonance occupation for a system containing a single planet with (a) $M_{pl} = M_{Jup}$, (b) $M_{pl} = 0.05 M_{Jup}$.  In each case, the planet has $e_{pl} = 0.1$ and $a_{pl} \sim 44.8$~AU.
Note the lower mass planet tends to trap particles in narrower resonances, closer to the planet's \sma.
}
\label{fig:synthcatmmr}
\end{center}
\end{figure}

Figure~\ref{fig:synthcatmmr} shows the resonance occupations for the
$e_{pl} = 0.1$ cases.  The effect of planetary mass is again highlighted,
with a high mass planet trapping particles in a few strong resonances, mostly
at higher semi-major axes, while a low mass planet has many weaker resonances, closer
to the planet's \sma.
More accurate simulated observations for specific systems and
telescopes are undertaken in Section~\ref{sec:compareprev}.

%The development of this synthetic catalogue has revealed several intriguing
%characteristics of planetary-disk systems \textit{\bf (SUCH AS!!)} which are not directly relevant to
%this paper. We reserve a more in-depth discussion of these features for a
%future paper.

\subsection{Dust Lifetime and Collisional Processes}
\label{subsec:collisionalproc}

As discussed in Section~\ref{subsec:recording}, our numerical procedure of using a small number of particles can lead to potential problems when terminating the simulation at a specified time, as some long-lived particles may still be active.  In a real debris disk, however, grain collisions would limit the maximum lifetime of particles.

\begin{figure}
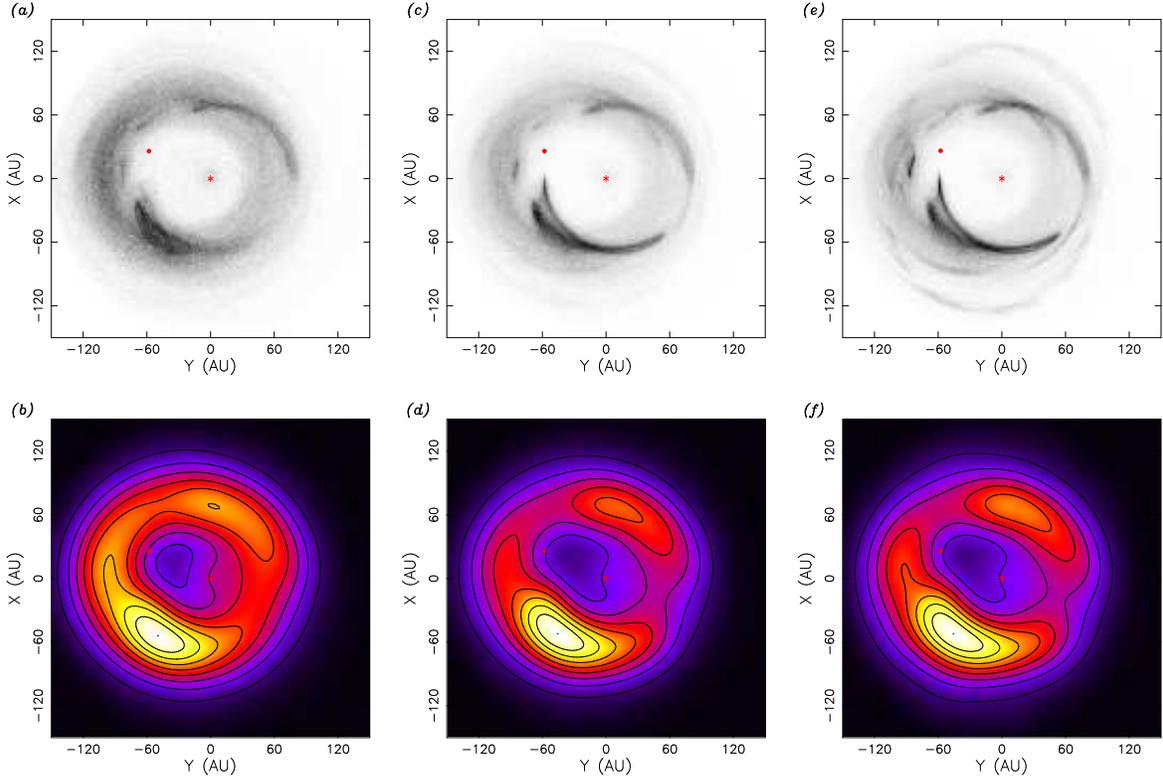

\begin{center}
\begin{tabular}{ccc}
\psfig{file=f6a_col.eps, width=0.3\textwidth, angle=270} &
\psfig{file=f6c_col.eps, width=0.3\textwidth, angle=270} &
\psfig{file=f6e_col.eps, width=0.3\textwidth, angle=270} \\
\psfig{file=f6b_col.eps, width=0.3\textwidth, angle=270} &
\psfig{file=f6d_col.eps, width=0.3\textwidth, angle=270} &
\psfig{file=f6f_col.eps, width=0.3\textwidth, angle=270}
\end{tabular}
\caption{Integrated dust distribution from a system containing a Jupiter mass planet with eccentricity 0.5 and \sma\ 44.8~AU, and 500 test particles with $\beta = 0.1$.  Parent bodies had initial orbital elements $1.4<a_{pb}/a_{pl}<2.0$, $0.0<e_{pb}<0.2$ and $0.0\degree<i_{pb}<8.0\degree$.  Simulated observations utilised a 2D Gaussian with FWHM of 40 AU.
Plots (a) and (b) show the dust distribution and a simulated observation after 11.7 million years (10\% of particles remaining).
Plots (c) and (d) show distribution and simulated observation after 40 million years (2\% of particles remaining).
Plots (e) and (f) show distribution and simulated observation after 153 million years (no particles remaining).
  See the electronic edition of the Journal for a colour version of this figure.
}
\label{fig:stoptime}
\end{center}
\end{figure}

In Figure~\ref{fig:stoptime} we show the results of a simulation of a Jupiter mass planet with $a_{pl} \sim 44.8$~AU and $e_{pl} = 0.5$ system containing 500 test particles with $\beta = 0.1$.  The test particles are released from parent bodies with $1.4<a_{pb}/a_{pl}<2.0$, $0.0<e_{pb}<0.2$ and $0.0\degree<i_{pb}<8.0\degree$.  The three panels show the dust distribution when the simulation is terminated after 11.7 million years, 40 million years and 153 million years, corresponding to 10\%, 2\% and 0\% of particles remaining active. It can be seen that
the basic disk structure is created rapidly, and the final few particles have a
limited impact despite having lifetimes many times the median.  The
longest-lived particles tend to `sharpen' the distribution and bring out
finer detail, which is hidden in our simulated observations (telescope FWHM 40 AU).  Thus,
the imposition of a cut-off lifetime (whether as a numerical convenience or as a substitute for dust destruction through collisions) is of little consequence as long as most ($>$90\%) particles have already been accreted or ejected.

%-------------------------------------------------------
\section{Modelling Observed Systems}
\label{sec:compareprev}

In this section we use the synthetic disk catalogue to select planetary configurations which generate structures similar to the observed debris disk systems Vega, \erid\ and Fomalhaut.  We then optimise the planetary and parent body parameters, and increase the resolution (by increasing the number of test particles) to produce more accurate synthetic observations, which we compare to previous observational and numerical studies of these three debris disk systems.

\subsection{Vega}
\label{subsec:vega}
Vega ($\alpha$ Lyrae) was the first star identified with an infrared
excess associated with a disk of dusty material \citep{aumann84} and has 
become the prototype for so-called `Vega-type' stars.  Subsequent observations 
with SCUBA \citep{holland98}
and the IRAM Plateau de Bure Interferometer \citep{whk02} confirm
that Vega is surrounded by a dusty disk, with two prominent clumps of
emission.  \citet{whk02} proposed that the twin lobes of emission
seen in these images are caused by dust trapped in MMRs with a
massive, eccentric planet ($M_{pl} = 3 M_{Jup}$, $e_{pl} = 0.6$)
orbiting Vega with a \sma\ of 40~AU.  In their model,
the initial test particle longitudes of perihelion were constrained to 
be aligned closely with the planet's longitude of perihelion (which is 
reasonable for a high eccentricity system).
They also noted that a wide variety of planetary parameters can produce
the twin-lobed structure seen in Vega.

Whilst our code was able to reproduce the results of \citet{whk02}, their
synthetic observations (at the resolution of the \citet{holland98} Vega 
observations) show nearly symmetrical emission, whereas the \citet{holland98} 
observations show a significant asymmetry.  
(It should be noted, however, that there are uncertainties in the
 \citet{holland98} observations.  In the model we are about to present,
 we are assuming that the observed asymmetry is real.)
Based on results from our
synthetic catalogue, we have modelled Vega using an entirely different
planetary configuration in an attempt to better match these observations.  
In our model, a more distant ($a_{pl} = 73.7$ AU), less eccentric 
($e_{pl} = 0.1$) 3 Jupiter mass planet reproduces the observed disk structure, 
with no constraints on the initial test particle perihelia (since we are 
modelling a lower eccentricity planet).  We use 5000 test particles 
released from parent bodies with initial orbital elements in the range 
$90<a_{pb}<120$, $0.0<e_{pb}<0.3$, and $0\degree<i_{pb}<8\degree$. 

\begin{figure}
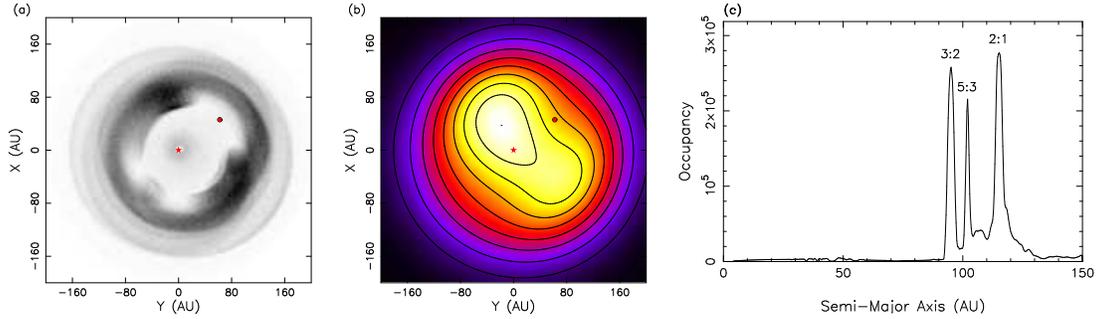

\centering
\begin{tabular}{ccc}
\psfig{file=f7a_col.eps,width=0.25\textwidth,angle=270} &
\psfig{file=f7b_col.eps,width=0.25\textwidth,angle=270} &
\psfig{file=f7c.eps,width=0.25\textwidth,angle=270}
\end{tabular}
\caption{
Results of Vega simulation using $a_{pl}=73.7$ AU, $e_{pl}=0.1$ and 5000 test particles with $\beta=0.05$ released from parent bodies with $90<a_{pb}<120$ AU, $0.0<e_{pb}<0.3$, and $0\degree<i_{pb}<8\degree$.  The simulation was terminated after $10^{8}$ years with 0.6\% of particles remaining.
  (a) Dust distribution -- note the two peaks as seen in the \citet{whk02} interferometric observations.
  (b) Simulated observation of the system using a resolution of 108 AU (equal to the resolution of the 850 \um\ SCUBA observations by \citealt{holland98}) and a 5 mJy solar flux, with contours at 2 mJy separation. The two peaks merge together at low resolution, although one dominates slightly.
  (c) Resonance occupancy in the Vega model.  See the electronic edition of the Journal for a colour version of this figure.
}
\label{fig:vega}
\end{figure}

\citet{dent00} state that to fit the observed spectral energy
distribution, the dust grains which make up the Vega disk must have
diameters in the range 60--400 $\mu$m, corresponding to $\beta$ = 0.02 --
0.11 for Vega's estimated mass and luminosity.  We have simulated this system 
with $\beta$ values of 0.02, 0.05 and 0.1, and found negligible differences 
between the synthetic observations.
(Note that \citet{whk02} used $\beta=0.01$ with $sw = 0$.)
Figure~\ref{fig:vega} shows the dust distribution, simulated observation and 
resonance occupancy plots obtained using our Vega model with $\beta=0.05$.

\begin{figure}
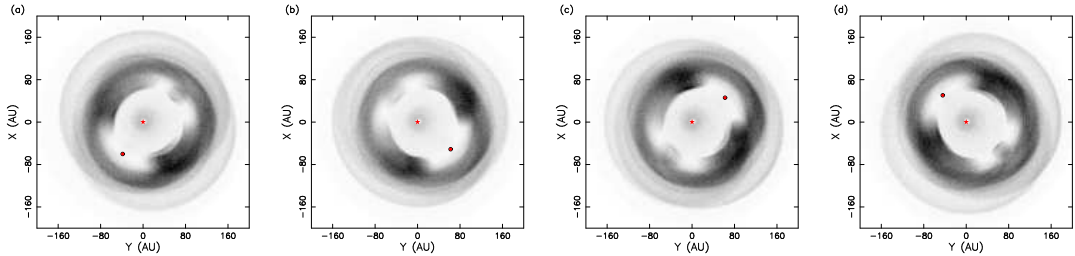

\centering
\begin{tabular}{cccc}
\psfig{file=f8a_col.eps,width=0.2\textwidth,angle=270} &
\psfig{file=f8b_col.eps,width=0.2\textwidth,angle=270} &
\psfig{file=f8c_col.eps,width=0.2\textwidth,angle=270} &
\psfig{file=f8d_col.eps,width=0.2\textwidth,angle=270}
\end{tabular}
\caption{Dust distribution generated by the Vega simulations (with $\beta=0.05$) at the four different recorded planetary phases.  The distribution rotates with the planet, meaning observations from Earth will show positional changes in the emission peaks over the planetary period.
}
\label{fig:vegaphase}
\end{figure}

We note that our model for Vega produces a dust distribution which is essentially stationary in the planet's frame of reference.  Such a distribution means that as the planet orbits the star, observations from Earth will show positional changes in the emission peaks over an orbital timescale.  Dust distributions from the four orbital phases recorded are shown in Figure~\ref{fig:vegaphase}.

The model accurately reproduces the twin lobes seen in the \citet{whk02}
observations, as well as the extended emission seen in
the lower resolution \citet{holland98} observations.  
Figure~\ref{fig:vega}a shows asymmetry in the two emission features, which 
are not collinear with the star nor the same distance from the star, 
as seen in the \citet{whk02} observations.
This model also provides reasonable constraints on the orbital parameters of 
the proposed planet; for example, the same model with $e_{pl}=0.2$ results
in a markedly different structure, as does a significantly less massive ($M_{pl} < M_{Jup}$) planet.  The primary concern with this model is the requirement 
that such a massive planet have formed or migrated to such a large distance 
from the central star, although the fact that Vega is estimated to be 2.5 
times as massive as the sun may mitigate this problem. Future detailed 
observations are required to test our planetary model.  

\subsection{\protect\erid}
\label{sebsec:epserid}
First identified as a Vega-type star in 1984 \citep{aumann85}, \erid\ has
been the subject of ongoing planetary speculation since disk images
taken at 850~\um\ by \citet{greaves98} showed a clumpy,
asymmetric ring, with a cleared region interior to $\sim 30$ AU.
Radial velocity observations by \citet{hatzes00} detected a $1.7 M_{Jup}$
planetary companion with  $a \sim 3.4$~AU and $e_{pl} \sim 0.6$.
Such a close companion, however, cannot account for the observed disk
structure seen by \citeauthor{greaves98}

\begin{figure}
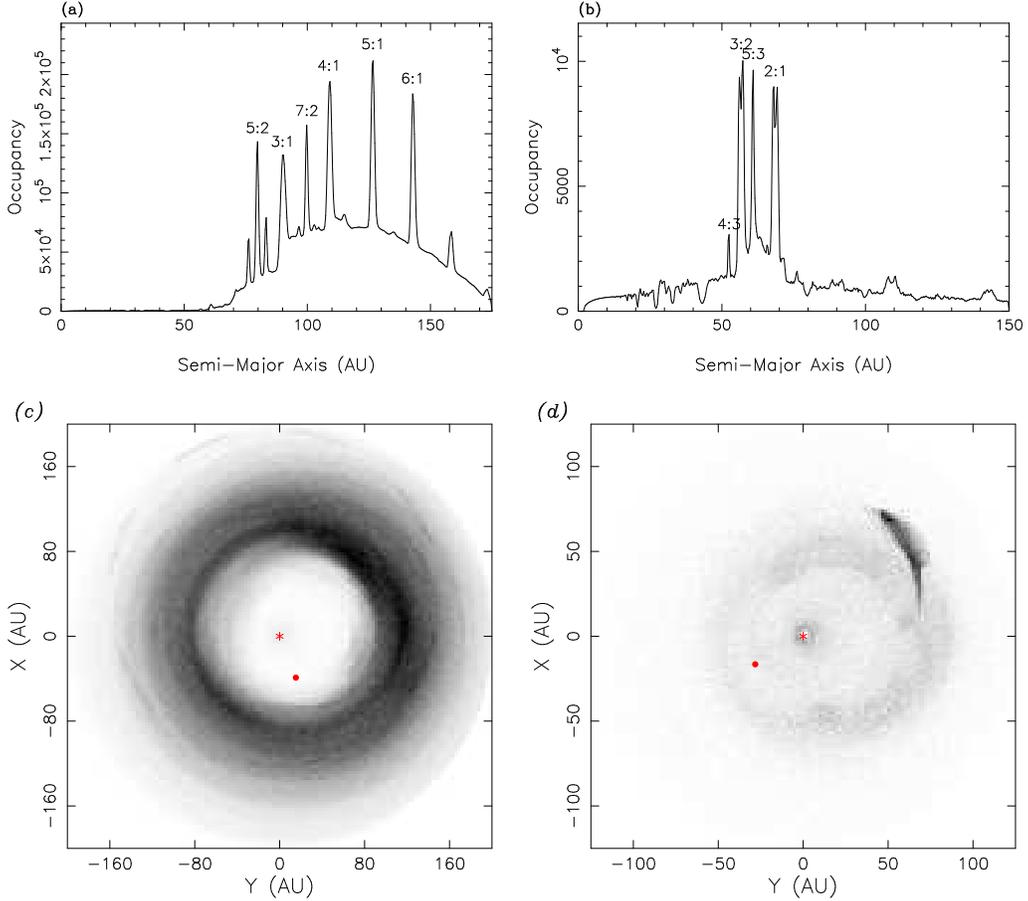

\centering
\begin{tabular}{cc}
\psfig{file=f9a.eps,width=0.3\textwidth,angle=270} &
\psfig{file=f9b.eps,width=0.3\textwidth, angle=270} \\
\psfig{file=f9c_col.eps,width=0.4\textwidth,angle=270} &
\psfig{file=f9d_col.eps,width=0.4\textwidth, angle=270}
\end{tabular}
\caption{Structures resulting from a Jupiter mass planet with $a_{pl} \sim 44.8$ AU and $e_{pl} = 0.3$.  Test particles have $\beta=0.1$, and are released from parent bodies with initial eccentricity range $e_{pb}=[0.0, 0.2]$.
(a) Resonance occupancy when the initial range of $a_{pb}$ is [2.0, 3.5]$a_{pl}$. The higher order n:1 resonances dominate.  Note the high number of total particle positions recorded, indicating longer average particle lifetimes.
(b) Resonance occupancy when the initial range of $a_{pb}$ is [1.0, 1.4]$a_{pl}$. The planet rapidly removes test particles, and only a few resonances close to the planet's \sma\ are populated.
(c) The dust distribution created when the initial range for $a_{pb}$ is [2.0, 3.5]$a_{pl}$.  A distinct ring structure is formed, containing two (unequal) density enhancements.
(d) The dust distribution created when the initial range for $a_{pb}$ is [1.0, 1.4]$a_{pl}$.  A ring structure is not generated}
\label{fig:bigplanets}
\end{figure}

Numerical simulations by \citet{oz00} have suggested the presence of a $M_{pl} = 0.2 M_{Jup}$, $a_{pl} = 55-65$ AU, $e_{pl} = 0.0$ planet, while \citet{qt02}
proposed a $M_{pl} = 0.1 M_{Jup}$, $a_{pl} = 41.6$ AU, $e_{pl} = 0.3$ planet in the \erid\ disk.
A planet with mass equal to or greater than Jupiter is
unlikely to be responsible for the observed disk structure, since massive
planets tend to trap particles
in the $n:1$ resonances \citep{kh03}.
Our simulations of a Jupiter mass planet, shown in
Figure~\ref{fig:bigplanets}a, demonstrate this effect.  For particles to occupy the $n+1:n$
resonances near a massive planet, their parent bodies must also reside
very close to the planet, and in this situation most particles are
quickly scattered or `leak past' the planet, as shown in
Figure~\ref{fig:bigplanets}b.  This results in a structure which
lacks the observed central cavity of \erid \, (see Figure~\ref{fig:bigplanets}c and \ref{fig:bigplanets}d).  The presence of the
observed central clearing would then require the gravitational influence of a second,
more massive planet at an intermediate \sma\ \citep{lz99}.

\begin{figure}
\centering
\begin{tabular}{cc}
\psfig{file=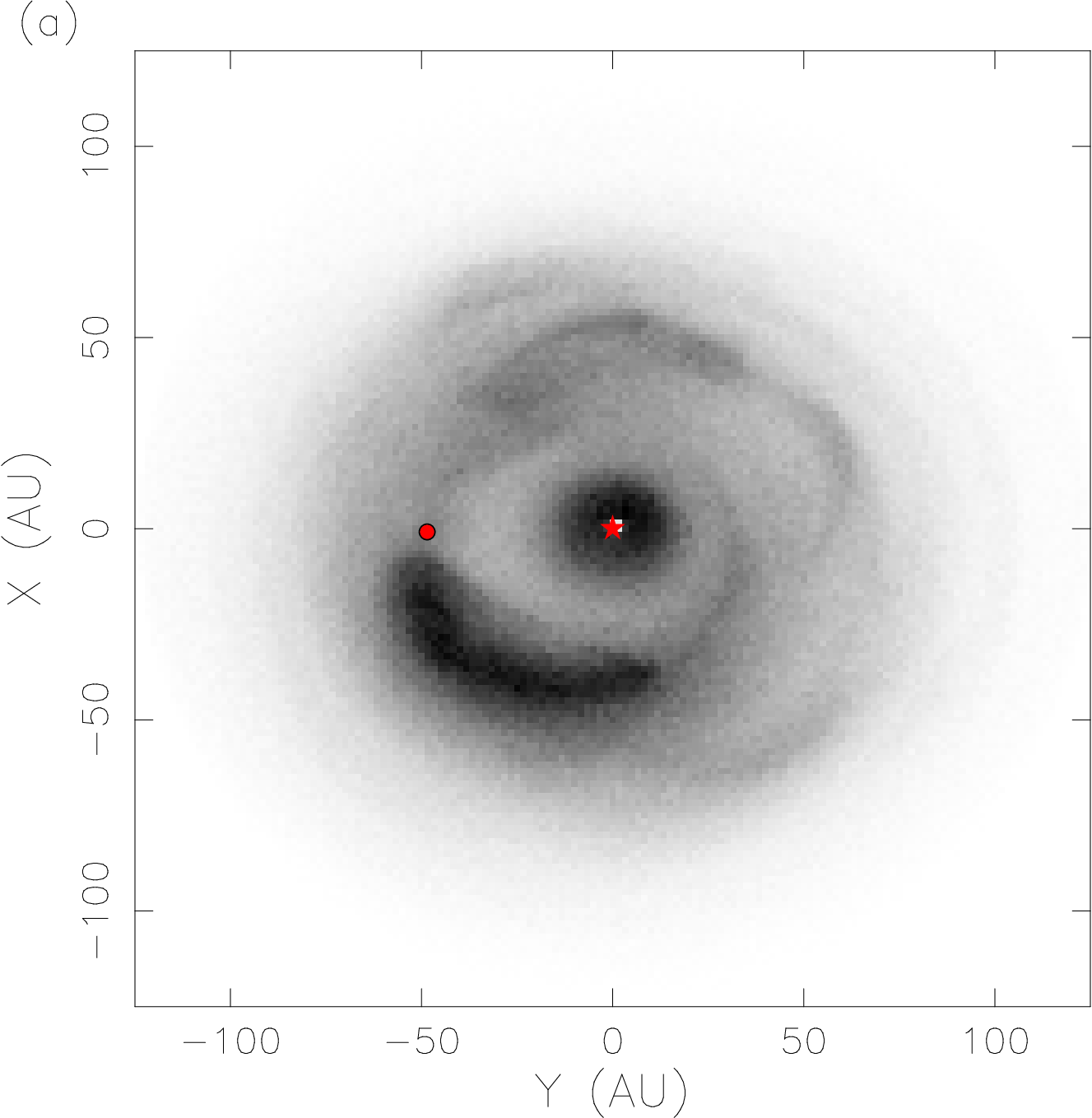,width=0.3\textwidth,angle=270} &
\psfig{file=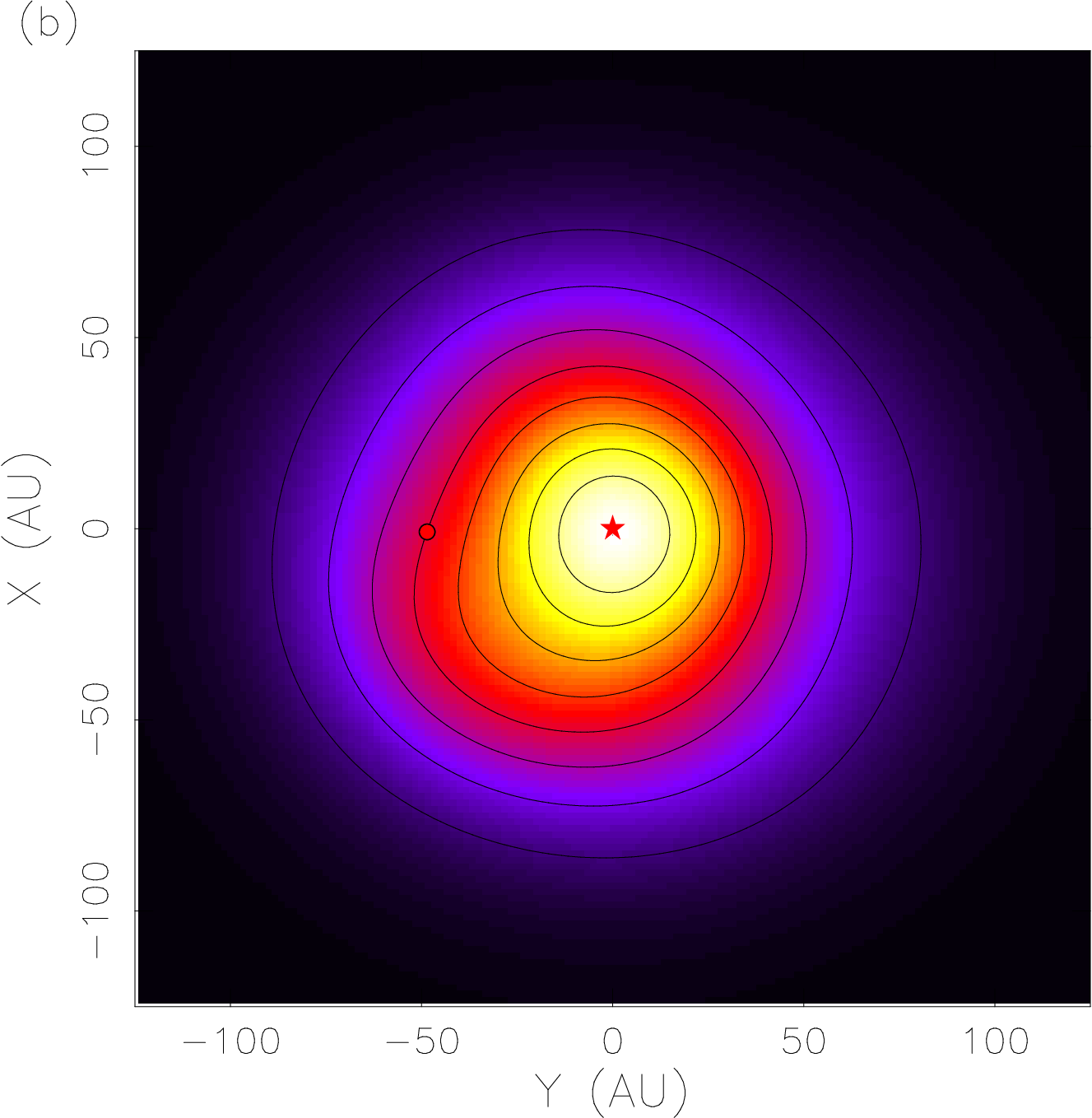,width=0.3\textwidth,angle=270} \\
\psfig{file=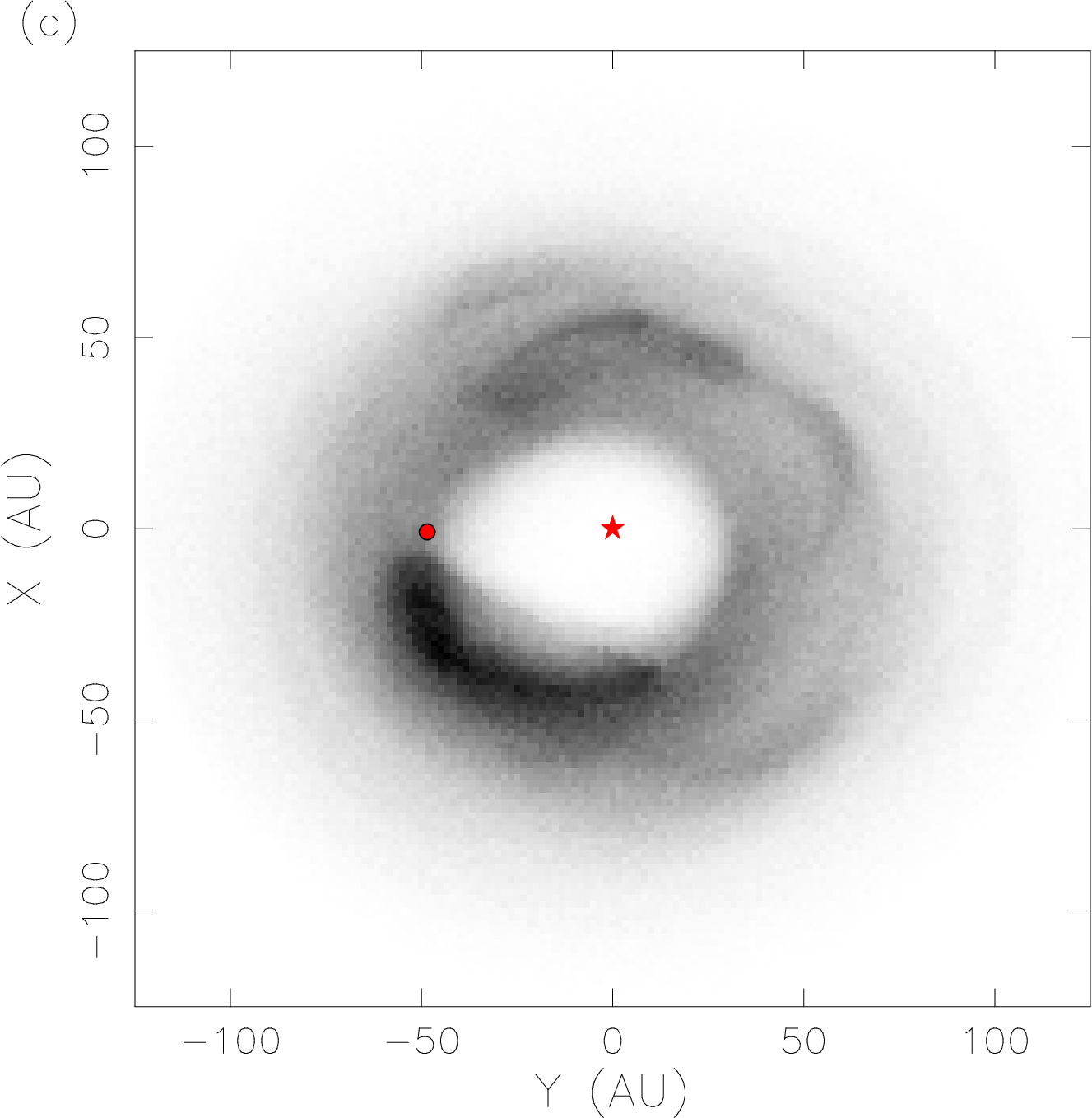,width=0.3\textwidth,angle=270} &
\psfig{file=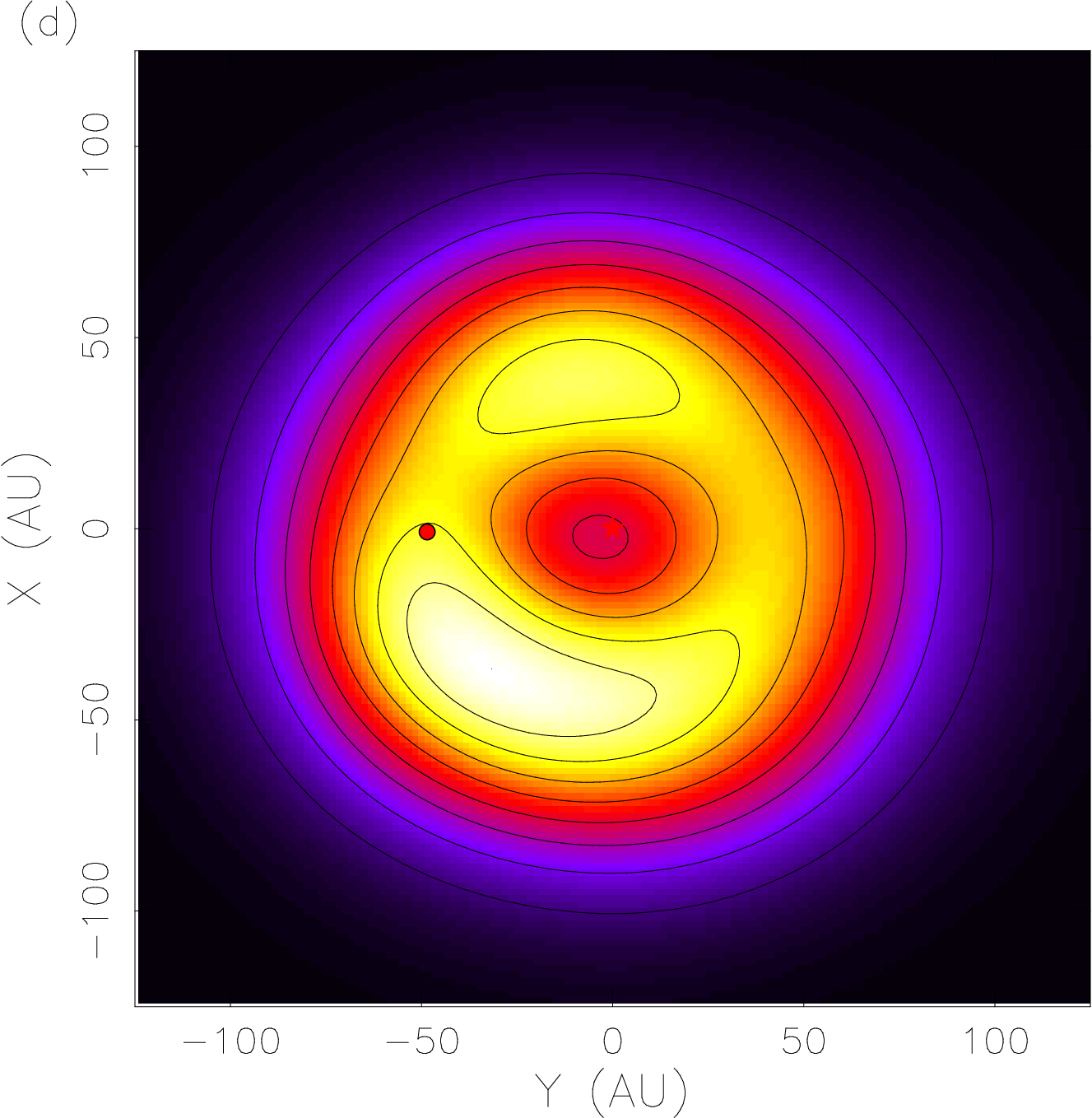,width=0.3\textwidth,angle=270} \\
\psfig{file=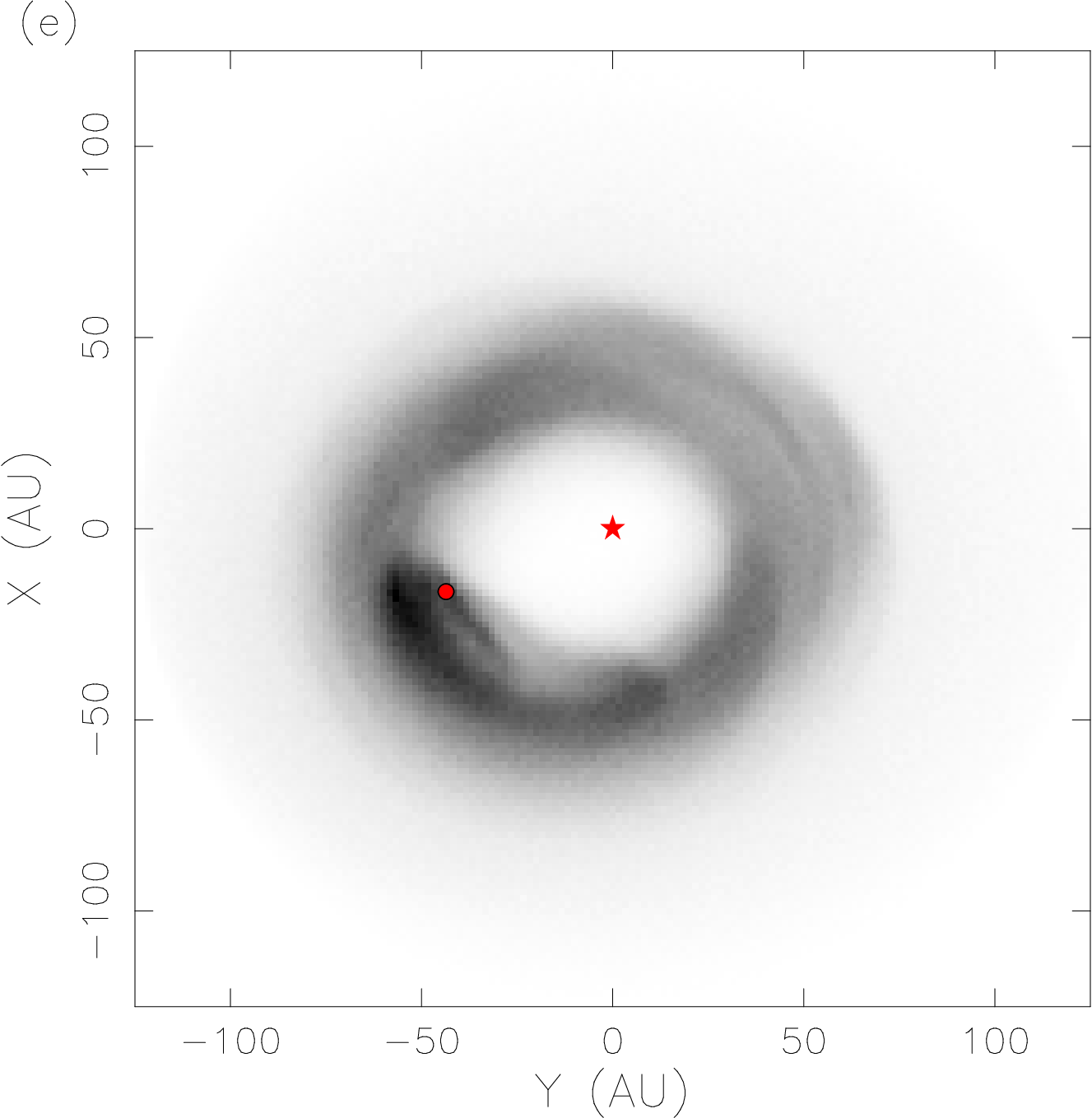,width=0.3\textwidth,angle=270} &
\psfig{file=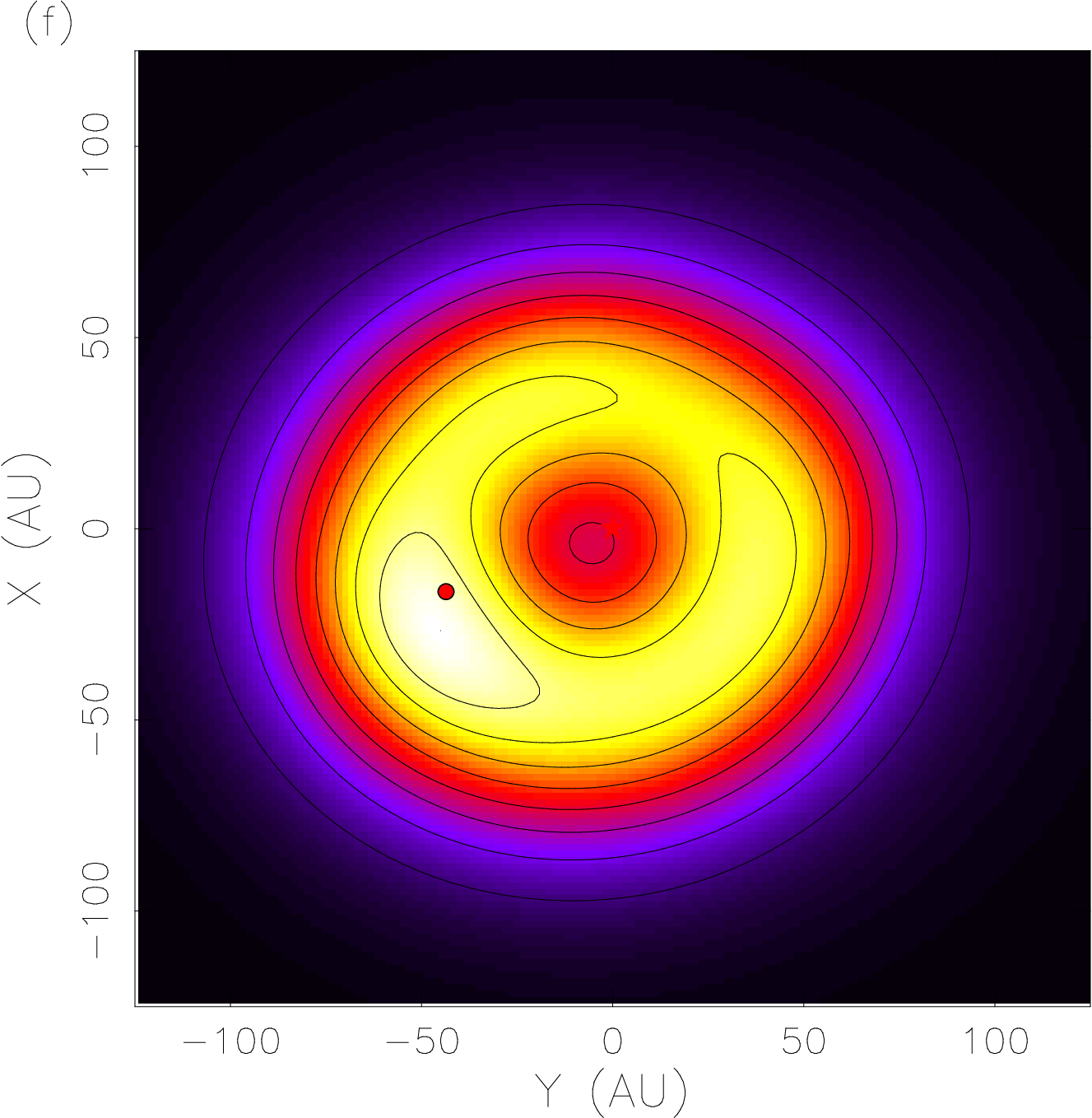,width=0.3\textwidth,angle=270}
\end{tabular}
\caption{Simulations of \protect\erid\ including a $M_{pl} = 0.1 M_{Jup}$, $e_{pl} = 0.3$ planet with a \sma\ of 41.6 AU, from a viewing angle of 30\degree. Parent bodies had $1.1<a_{pb}/a_{pl}<1.5$, $0<e_{pb}<0.3$ and $0.0<i_{pb}<8.0$.  The telescope beam FWHM was set to 45 AU - the approximate beam diameter of the \citet{greaves98} images.
(a) Dust distribution, \protect$\beta=0.1$, all particles allowed.
(b) Simulated observation, \protect$\beta=0.1$, all particles allowed. Emission close to the star dominates and the ring structure is faint.
(c) Dust distribution, \protect$\beta=0.1$, particles with \smas\ less than the planet removed.
(d) Simulated observation, \protect$\beta=0.1$, particles with \smas\ less than the planet removed.
(e) Dust distribution, \protect$\beta=0.01$, particles with \smas\ less than the planet removed.
(f) Simulated observation, \protect$\beta=0.01$, particles with \smas\ less than the planet removed.
  See the electronic edition of the Journal for a colour version of this figure.
}
\label{fig:eridani}
\end{figure}

\citeauthor{oz00} consider only particles in the 2:1 and 3:2 resonances
(in equal proportions), while \citeauthor{qt02} consider only
particles with $a_{tp} = [1.1, 1.5]a_{pl}$.  An examination of our
results shows why; if particles with \smas\ smaller than the \sma\ of
the planet are allowed (as in Figure~\ref{fig:eridani}a), emission
from particles close to the star dominates and a ring structure is not
seen (see Figure~\ref{fig:eridani}b). The implied ejection of particles from the inner regions of the disk
hints at another undiscovered and more massive planet at an intermediate
distance from the star.  The requirements for such a planet are discussed
below.

We suggest that the system proposed by \citeauthor{oz00} is unlikely to
be responsible for the \erid\ disk for two reasons: (1) their model
is symmetrical whereas the original observations by \citet{greaves98}
and subsequent 350 \um\ observations by D. J. Wilner (2004, private
communication) state that only the most prominent clump (in the southeast of the image) is definitely
present, and (2) our simulations show that dust released in a range
encompassing the 2:1 and 3:2 resonances will always occupy other
resonances such as the 5:3, and that the resonance occupancies are rarely in equal
proportion.  We therefore investigate the model proposed by
\citeauthor{qt02} in more detail.

Our models differ somewhat from those of \citeauthor{qt02} in that
we use 5000 test particles (increasing the resolution by an order of
magnitude), we include solar wind drag, and model the system in three
dimensions.  The simulations were terminated after $3\times10^{8}$
years, when less than 1\% of the test particles remained active.  Because the
dominant grain size in the \erid\ disk has not been well determined, we
have simulated the system using two $\beta$ values, $\beta=0.1$ and 0.01.
Initial parent body \smas\ and eccentricities are set in the range
[1.1 $a_{pl}$, 1.5 $a_{pl}$] and [0,0.4] respectively, while initial
inclinations are in the range [0\degree,8\degree]. Our results
are shown in Figure~\ref{fig:eridani}.

\begin{figure}
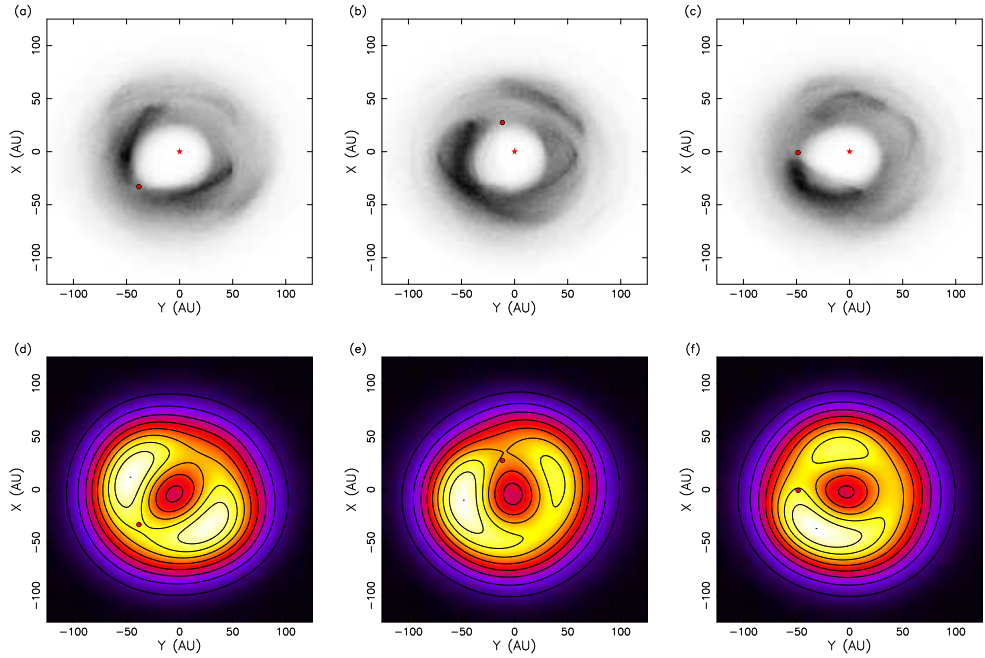

\centering
\begin{tabular}{ccc}
\psfig{file=f11a_col.eps,width=0.25\textwidth,angle=270} &
\psfig{file=f11b_col.eps,width=0.25\textwidth,angle=270} &
\psfig{file=f11c_col.eps,width=0.25\textwidth,angle=270} \\
\psfig{file=f11d_col.eps,width=0.25\textwidth,angle=270} &
\psfig{file=f11e_col.eps,width=0.25\textwidth,angle=270} &
\psfig{file=f11f_col.eps,width=0.25\textwidth,angle=270}
\end{tabular}
\caption{Dust distribution and synthetic observations from by the \erid\ simulations (with $\beta=0.1$ and particles with \smas\ interior to the planet removed) at three different planetary phases.  Dust distributions are shown in (a) to (c), with the corresponding synthetic observations shown in (d) to (f).  The distribution does not rotate with the planet, but the emission peaks show positional changes over the course of an orbit.  See the electronic edition of the Journal for a colour version of this figure.
}
\label{fig:eridphase}
\end{figure}

Both the $\beta = 0.1$ and $\beta = 0.01$ models reproduce the ring
structure seen in the sub-mm observations, possessing a single
prominent emission maxima and one or more secondary maxima.  As
expected, the contrast and detail is higher in the $\beta = 0.01$
case, since more particles are trapped in resonances for longer.
Unlike Vega, this dust distribution generated by this model is not fixed in the planet's reference frame, but the emission peaks do show positional changes over a planetary orbit, as shown in  Figure~\ref{fig:eridphase}.  Our
results confirm the model proposed by \citet{qt02} as a
viable explanation for the \erid\ system, with the caveat that an
additional massive body must exist interior to the dust-sculpting
planet's orbit.

In order to confirm that an inner planet is responsible for clearing the inner regions of the disk, one would like to simply introduce a second planet into the \erid\ model.  However, since the introduction of a second planet introduces variations into
the orbital elements of the primary outer planet, recording particle
positions with the outer planet in an fixed position over time is
no longer possible.  We can instead introduce an inner planet to our model at a range of \smas\ and make use of test particle occupancy versus \sma\ plots to determine if the inner planet effectively removes particles from the inner $\sim 40$ AU of the system.

\begin{figure}
\centering
\begin{tabular}{cc}
\psfig{file=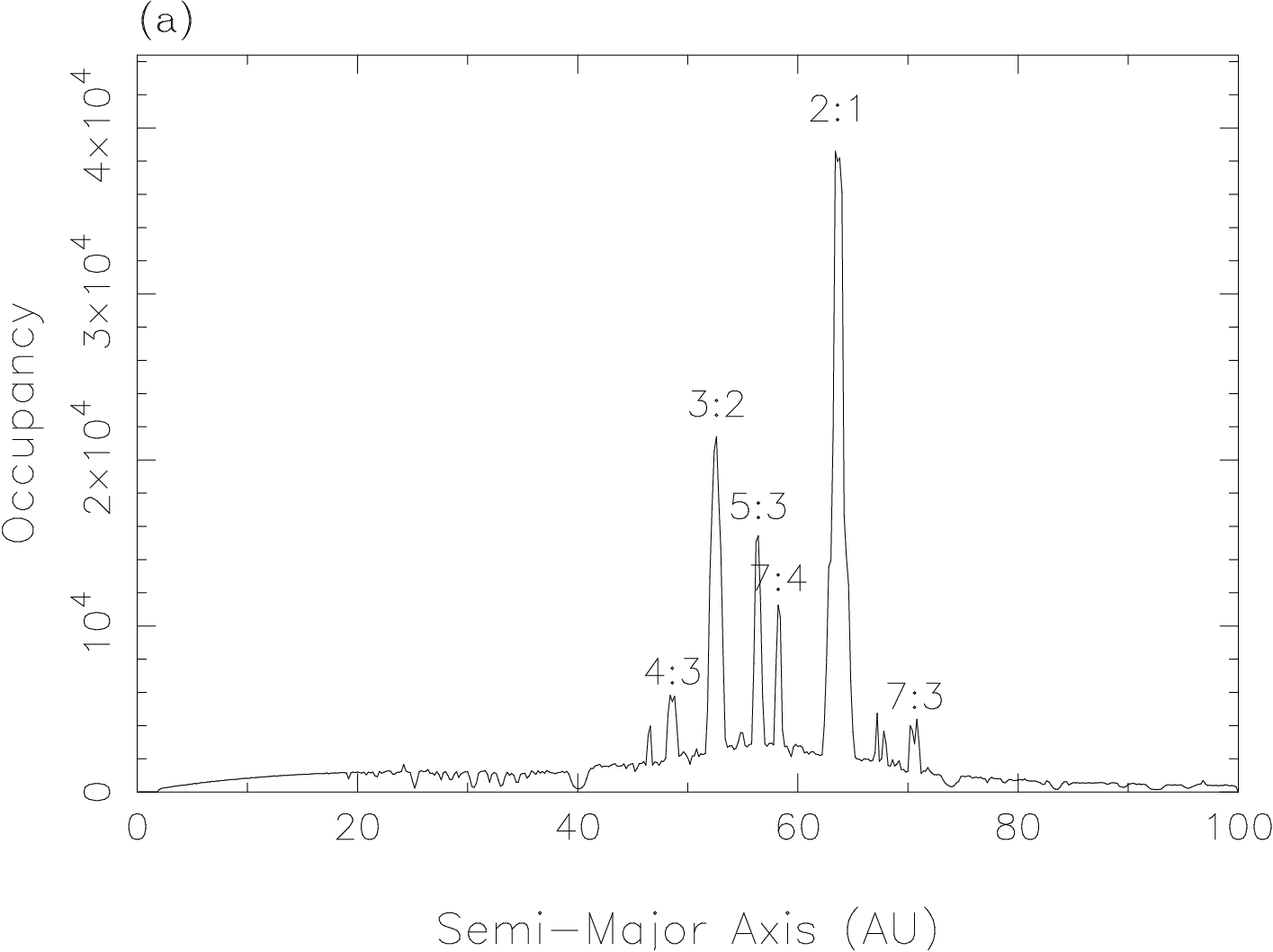,width=0.3\textwidth,angle=270} &
\psfig{file=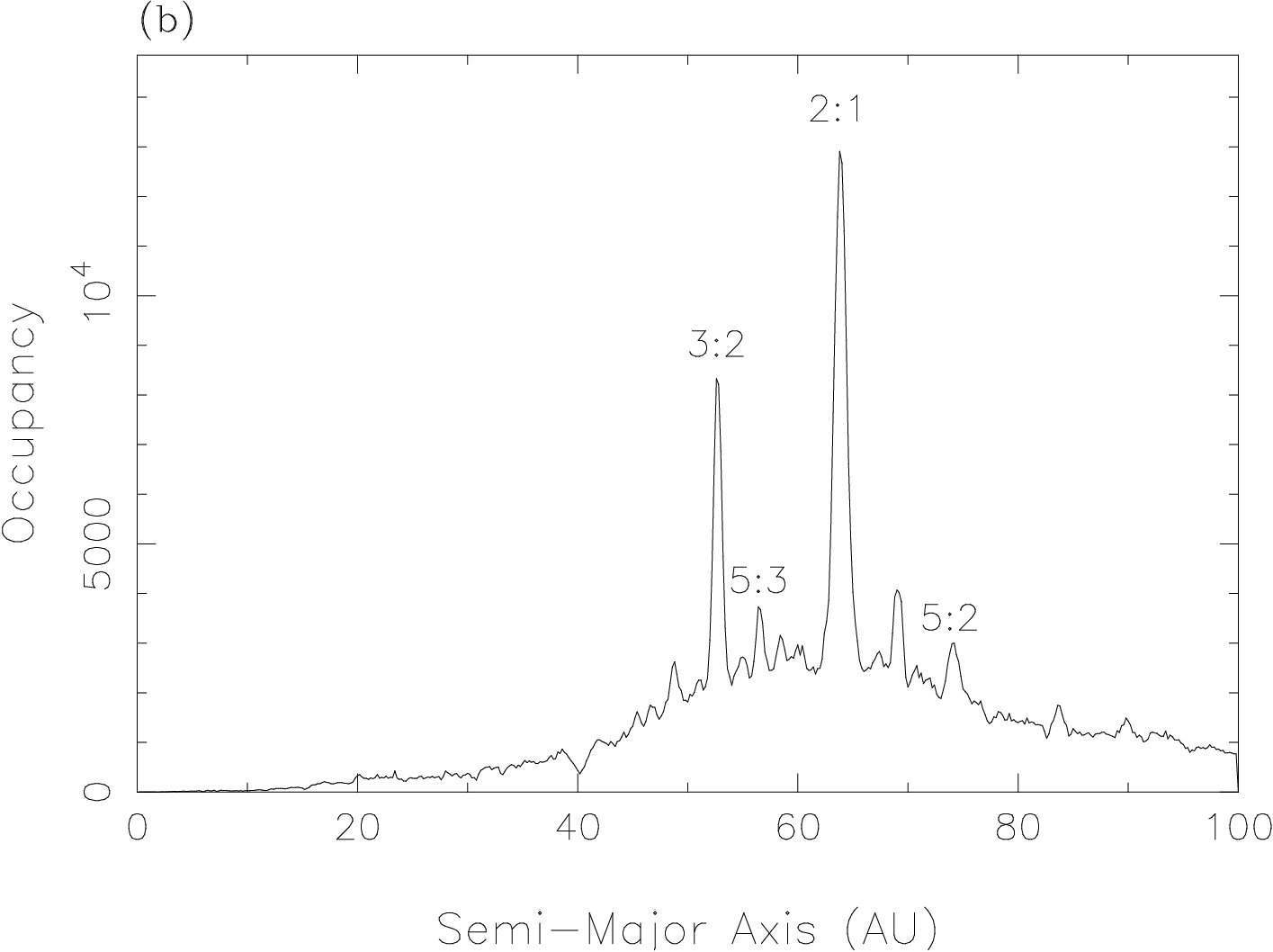,width=0.3\textwidth,angle=270} \\
\psfig{file=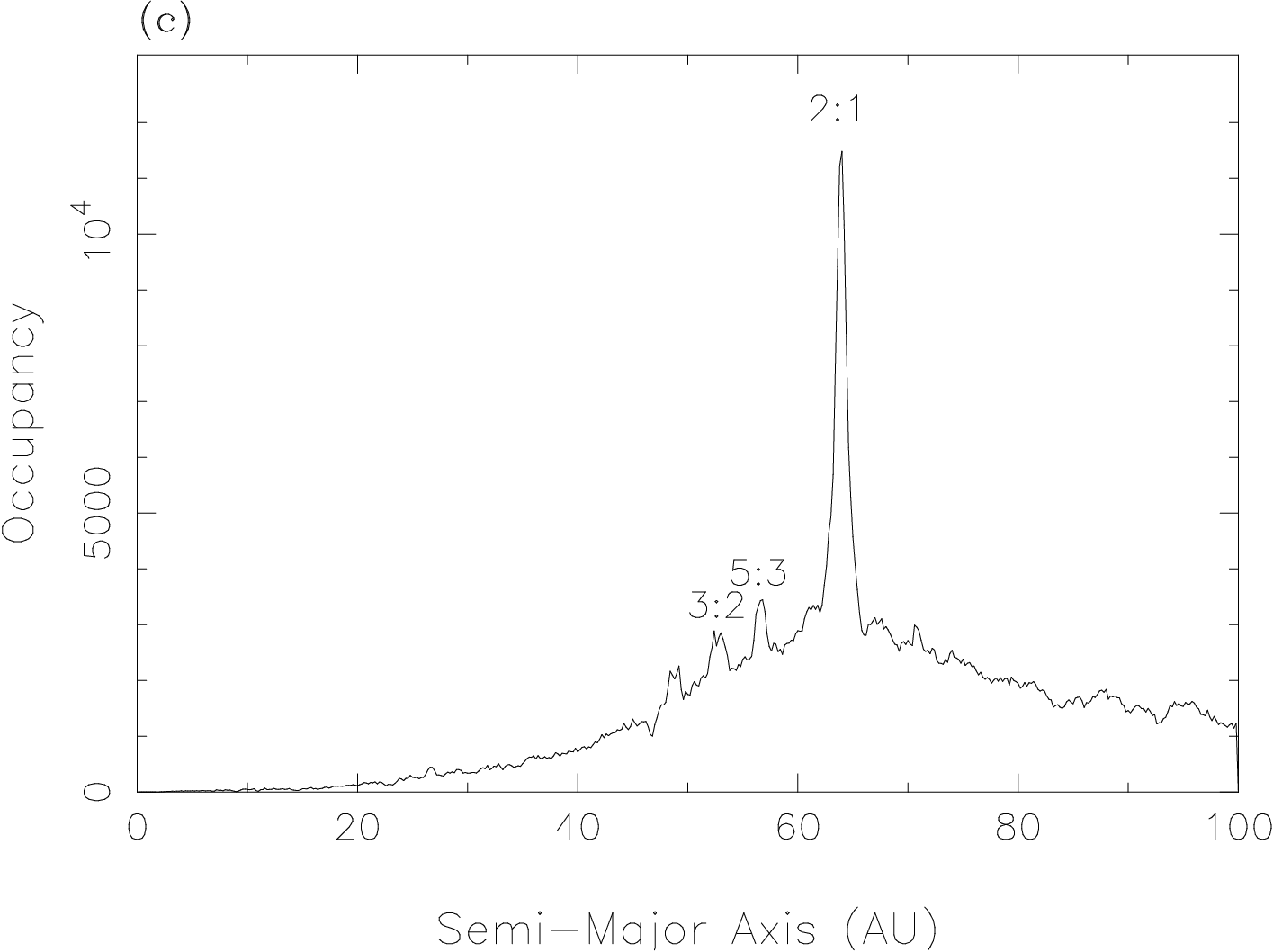,width=0.3\textwidth,angle=270} &
\psfig{file=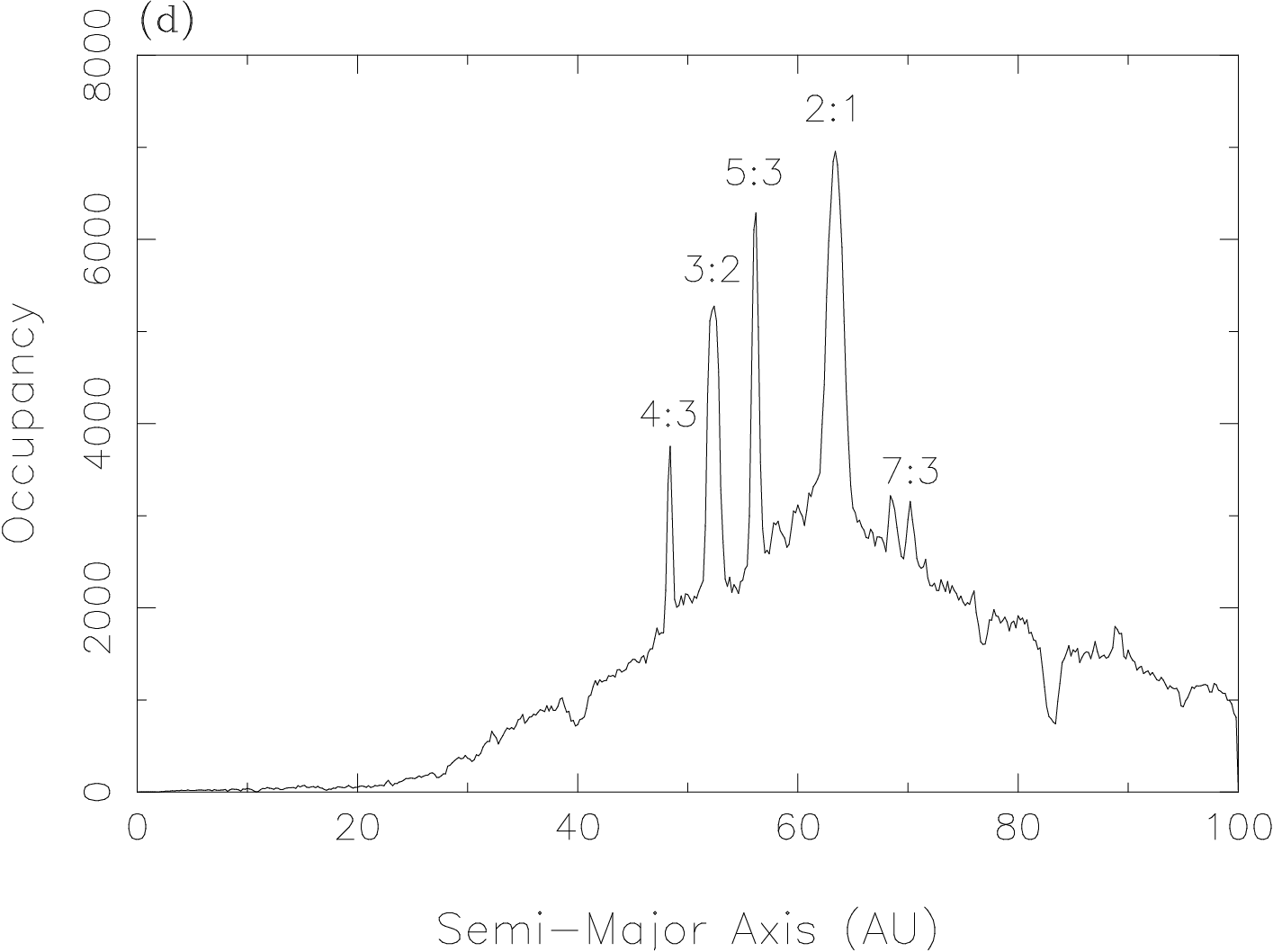,width=0.3\textwidth,angle=270}
\end{tabular}
\caption{Occupancy versus \sma\ plots showing the effect of adding a Jupiter mass, $e_{pl} = 0.3$ inner planet to the \erid\ model.  Plot (a) shows the plot for the unmodified \erid\ model - note test particles occupancy of \smas\ interior to 40 AU.  Plots (b), (c) and (d) show the results for systems with an inner planet added with a \sma\ of 10, 15 and 18 AU respectively.  In each case, test particle occupancy of \smas\ interior to 40 AU is substantially reduced (note the different scales on the four graphs).  Simulations contained five hundred $\beta=0.1$ test particles were released from parent bodies with $46<a_{pb}<62$ AU, $0.0<e_{pb}<0.4$ and $0.0\degree<i_{pb}<8.0\degree$ and were terminated after $10^{8}$ years.
}
\label{fig:twoplanet}
\end{figure}

Plots of occupancy versus \sma\ are shown in Figure~\ref{fig:twoplanet} for our preferred model of \erid\ (Figure ~\ref{fig:twoplanet}a) and models including the addition of a second, Jupiter mass planet with $e_{pl} = 0.3$ at three different \smas; $a_{pl}$ = 10, 15 and 18 AU (Figure~\ref{fig:twoplanet}b, \ref{fig:twoplanet}c, and \ref{fig:twoplanet}d).
We see that the addition of a Jupiter mass planet with a \sma\ between $10-18$ AU is capable of substantially
reducing particle occupation of \smas\ less than that of the exterior planet.
We found that a planet located too far from the star ($a_{pl} \geq 20$ AU)
disrupted the orbit of the outer planet and resulted in a dynamically unstable system.
Note the change in the scaling of the graphs in Figure~\ref{fig:twoplanet}. 
We can clearly see that when an
inner planet is included, the resonance occupancy decreases, reflecting a reduction in particle lifetimes. An understanding of the exact effects of a massive inner planet will have
to await detailed simulations with large particle numbers, which can
resolve the disk at all times.  Such an investigation is planned for a future paper.

\subsection{Fomalhaut}
\label{sec:fomalhaut}
Unlike the previous systems, Fomalhaut has not yet been subject to
detailed numerical simulations.  The Fomalhaut system also differs
markedly in that 850 \um\ SCUBA observations suggest the system is seen nearly
edge-on and consists of a dusty torus, rather than a thin disk
\citep{holland98}. The improved spatial
resolution offered by the 450 \um\ SCUBA images of \citet{holland03} confirm the torus
structure, while revealing a previously unseen arc of emission near or
within the torus.  This departure from a uniform structure is
strongly suggestive of a planetary presence.

\begin{figure}
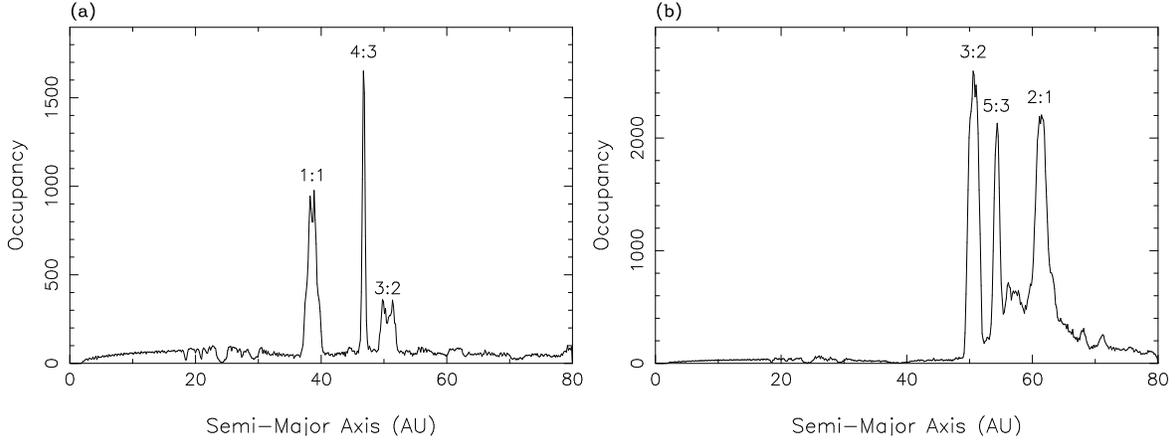

\centering
\begin{tabular}{cc}
\psfig{file=f13a.eps,width=0.35\textwidth,angle=270}
\psfig{file=f13b.eps,width=0.35\textwidth,angle=270}
\end{tabular}
\caption{The resonance occupancy for a 500 particle simulation of a system containing a Jupiter mass planet with $a_{pl}=40$ AU and $e_{pl}=0.1$.  Test particles had $\beta = 0.1$.
(a) With parent bodies in the range $35<a_{pb}<45$ AU, $0.0<e_{pb}<0.3$ and $0.0\degree<i_{pb}<8.0\degree$, the 1:1 resonance is prominent.
(b) When parent bodies are in the range $45<a_{pb}<60$ AU, $0.0<e_{pb}<0.3$ and $0.0\degree<i_{pb}<8.0\degree$, there is no population of the 1:1 resonance.
  }
\label{fig:11resonance}
\end{figure}

The generation of a single arc of emission can be achieved by a 1:1
resonance, as suggested by \citet{holland03}.
%% using results from Ozernoy \etal (2000).
Inspection of results for resonance occupancy obtained in the construction of our synthetic catalogue,
however, indicate that the 1:1 resonance is difficult to populate under normal circumstances. An example is shown in Figure~\ref{fig:11resonance}, where a Jupiter mass planet populates the 1:1 resonance when parent bodies exist interior to the planet, but does not populate the 1:1 resonance when the parent bodies are all external to the planet.  We consider it unlikely that parent bodies with \smas\ slightly less than the planet's \sma\ would survive for the required length of time.  As an alternative explanation of the single arc feature, we find that systems including a massive ($M_{pl} \geq M_{Jup}$), moderately eccentric ($0.3 \leq e_{pl} \leq 0.5$)
planet can induce a single emission arc over a background ring by
trapping many particles in the $n:1$ resonances, where $n>1$.

\begin{figure}
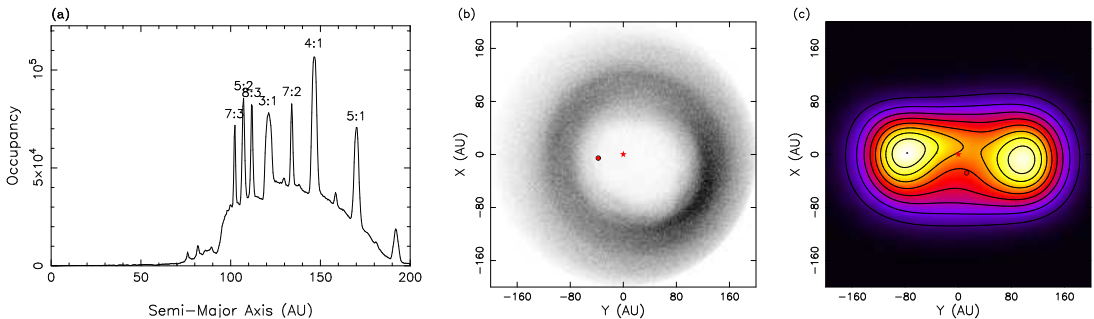

\centering
\begin{tabular}{ccc}
\psfig{file=f14a.eps,width=0.25\textwidth,angle=270} &
\psfig{file=f14b_col.eps,width=0.25\textwidth,angle=270} &
\psfig{file=f14c_col.eps,width=0.25\textwidth,angle=270}
\end{tabular}
\caption{Simulations of the Fomalhaut system using a 2 $M_{Jup}$
  planet with $e_{pl}=0.4$ and $a_{pl} \sim 59$ AU.
  (a) The resonance occupation - note that the n:1 resonances dominate.
  (b) Face-on distribution of particles.
  (c) Simulated observation of system at 450\um\ from an inclination of 70\degree.  The telescope FWHM of is 50 AU, equivalent to the resolution of SCUBA at 450$\mu$m. Contours show equal intensity increments.  See the electronic edition of the Journal for a colour version of this figure.
  }
\label{fig:fomalhaut}
\end{figure}

After examining our synthetic catalogue, we chose a model with the
following parameters: $M_{\star} = 2.3 M_{\odot}$ \citep[appropriate for an A3V star such as Fomalhaut -- ][]{bar97}, $M_{pl} = 2 M_{Jup}$, $e_{pl}=0.4$, and $a_{pl}
\sim 59$~AU. Parent bodies were distributed with initial orbital
parameters in the following ranges: $0<e_{pb}<0.3$,
$100<a_{pb}<180$~AU, $0<i_{pb}<25$\degree.  The higher parent body inclinations naturally generates a torus structure, which is believed to exist in the Fomalhaut system.  The simulation used 1000
test particles with $\beta = 0.05$, since the mass of dust grains
in the Fomalhaut system with diameters $>$ 100\um\ (corresponding to
values of $\beta > 0.1$) is believed to be negligible \citep{dent00}.
The results of our simulations, which ran until no particles remained
after 212 million years, are shown in Figure~\ref{fig:fomalhaut}.
The simulated observations bear a close resemblance to the `raw' 450\um\ image
of Fomalhaut presented in \citet{holland03}.

\begin{figure}
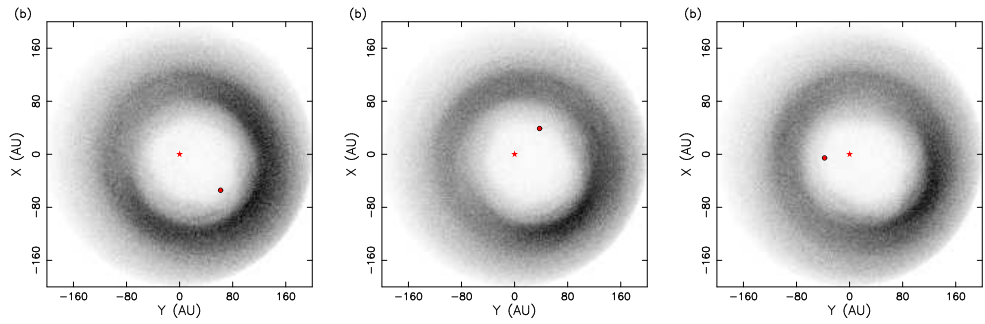

\centering
\begin{tabular}{ccc}
\psfig{file=f15a_col.eps,width=0.25\textwidth,angle=270} &
\psfig{file=f15b_col.eps,width=0.25\textwidth,angle=270} &
\psfig{file=f15c_col.eps,width=0.25\textwidth,angle=270}
\end{tabular}
\caption{Dust distribution generated by the Fomalhaut simulations at three different planetary phases.  The distribution does not rotate with the planet, and remains relatively constant in a fixed reference frame.
}
\label{fig:fomalphase}
\end{figure}

This planetary configuration does generates a dust distribution which does not rotate with the planet, but appears fixed from a viewpoint external to the system over an orbital period.  Thus, observations of Fomalhaut over time would show no change in emission if this model is correct.  The effects of planetary phase are shown in Figure~\ref{fig:fomalphase}.

Whilst the planetary configuration we have presented here displays a close similarity to the
observations of Fomalhaut to date, as noted above results from our synthetic catalogue suggest that several parameters from the model may be varied slightly whilst still producing similar results.  The
\sma\ of the planet is well constrained by the location of the
dust ring, but the the production of a single arc of emission is seen with several combinations of massive, moderate eccentricity planets.  Thus, this model is only representative of a class of planetary configurations which could be responsible for the structure seen in the Fomalhaut disk.

%-------------------------------------------------------
\section{Conclusions}
\label{sec:conclusions}
We have modified an N-body symplectic integrator to include the effects of \pr\ and solar wind drag in order to model collionless debris
disks, and used the modified integrator to produce a synthetic disk
catalogue containing approximately 300 model disks.  The catalogue
has two applications: the comparison of theoretical predictions of disk structure to numerical results, and to give an idea of the parameter space which might
satisfy a particular observed system, possibly identifying planets which are presently undetectable through Doppler shift techniques. Using this synthetic catalogue as a guide, we have produced `best fit' models for three observed debris disk systems whose disk structures can be explained by the presence of a planetary companion or companions.

We have modelled Vega using a planet with $M_{pl} = 3 M_{Jup}$, $a_{pl} = 73.7$ AU and $e_{pl} = 0.1$.  The dust distribution generated by this model rotates with the planet, meaning future observations of Vega should show changes in spatial dust emission as the planetary phase changes.  However, since a range of planetary parameters can produce the dust distribution inferred for the Vega system, our solution is representative of that produced by a class of massive ($M_{pl} > M_{Jup}$), low eccentricity ($e_{pl} < 0.2$) planetary companions.

\erid\ was modelled using a planet with $M_{pl} = 0.1 M_{Jup}$, $a_{pl} = 41.6$ AU and $e_{pl} = 0.3$, a model first proposed by \citet{qt02}.  Although the dust distribution generated by this model does not rotate with the planet, it does change over the course of an orbital period, offering hope for confirmation by future observations.  However, this model is dependent on the existence of a second, massive inner planet to clear particles interior to the modelled planet, which cannot be simulated using the techniques outlined in this paper.

We modelled Fomalhaut using a planet with $M_{pl} = 2 M_{Jup}$, $a_{pl} = 59 AU$ and $e_{pl}=0.4$.  The dust distribution generated by this model does not change significantly over the course of a planetary orbit, and thus the effect of planetary phase would be difficult to detect in future observations.  Like the model presented for Vega, our model for Fomalhaut is the best match to current observations from a class of massive, moderate eccentricity planets which generate similar structure.

Numerical modelling of planets in dusty debris disks has shown that planets which are undetectable through present techniques are likely to be responsible for the observed asymmetries of known debris disk systems.  It should be noted, however, that it is difficult to positively identify a unique planetary companion without temporal information showing the changes in disk structure over the course of a planetary orbit.

%-------------------------------------------------------
\acknowledgments
The authors wish to thanks Hal Levison for useful discussions about
RMVS3 and Dave Wilner for providing an unpublished 350 \um\
\erid\ image. We also thank the anonymous referee for providing useful
feedback and suggestions.
AD was supported by a Summer Vacation Scholarship from
the Swinburne Centre for Astrophysics and Supercomputing.  All
simulations were run on the Swinburne supercomputer\footnote{\tt http://supercomputing.swin.edu.au/}.

Our synthetic debris disk catalogue is available online at:\\
{\tt http://astronomy.swin.edu.au/debrisdisks/}

%-------------------------------------------------------


\begin{thebibliography}{}
\bibitem[Aumann \etal(1984)]{aumann84} Aumann, H. H. \etal\ 1984, \apj, 278, L23
\bibitem[Aumann (1985)]{aumann85} Aumann, H. H. 1984, PASP, 97, 885
\bibitem[Backman \& Paresce (1993)]{backman93} Backman, D. E. \& Paresce, F. 1993, 
in Protostars and Planets III,  ed. E.H. Levy \& J.I. Lunine (University of Arizona Press), 1253
\bibitem[Backman \etal(1995)]{backman95} Backman, D. E., Dasgupta, A. \& Stencel, R. E. 1995, 
\apj, 450, L35
\bibitem[Barrado y Navascues \etal(1997)]{bar97} Barrado y Navascues, D., Stauffer, J. R., Hartmann, L. \& Balachandran, S. C. 1997, \apj, 475, 313
\bibitem[Burns \etal(1979)]{burns79} Burns, J. A., Lamy, P. L. \& Soter, S. 1979, \icarus, 40, 1
\bibitem[Clampin \etal(2003)]{clampin03} Clampin, M. \etal\ 2003, \aj, 126, 385
\bibitem[Dent \etal(2000)]{dent00} Dent, W. R. F., Walker, H. J., Holland, W. S. \& Greaves, J. S. 2000, \mnras, 314, 702
\bibitem[Dermott \etal(1994)]{dermott94} Dermott, S. F., Jayamaran, S., Xu, Y. L., Gustafson, B. A. S. \& Liou, J.-C. 1994, \nat, 369, 719
\bibitem[Duncan \etal(1998)]{dll98} Duncan, M. J., Levison, H. F. \& Lee, M. H. 1998, \aj, 116, 2067
\bibitem[Greaves \etal(1998)]{greaves98} Greaves, J. S. \etal\ 1998, \apj, 506, L133
%\bibitem[Gustafson(1994)]{guf94} Gustafson, B. A. S. 1994, \areps, 22, 553
\bibitem[Hatzes \etal(2000)]{hatzes00} Hatzes, A. P. \etal\ 2000, \apj, 544, L145
\bibitem[Holland \etal(1998)]{holland98} Holland, W. S. \etal\ 1998, \nat, 392, 788
\bibitem[Holland \etal(2003)]{holland03} Holland, W. S. \etal\ 2003, \apj, 582, 1141
\bibitem[Kuchner \& Holman(2003)]{kh03} Kuchner, M. J. \& Holman, M. J. 2003, \apj, 588, 1110
\bibitem[Landgraf \etal(2002)]{land02} Landgraf, M., Liou, J.-C., Zook, H. A. \& Gr\"{u}n, E. 2002, \aj, 123, 2857
\bibitem[Levison \& Duncan(1994)]{ld94} Levison, H. F. \& Duncan M. J. 1994, \icarus, 108, 18
\bibitem[Liou \& Zook(1997)]{lz97} Liou, J.-C. \& Zook, H. A. 1997, \icarus, 128, 354
\bibitem[Liou \& Zook(1999)]{lz99} Liou, J.-C. \& Zook, H. A. 1999, \aj, 118, 580
\bibitem[Liseau(1999)]{liseau99} Liseau, R. 1999, \aap, 348, 133
\bibitem[Moro-Martin \& Malhotra(2002)]{mm02} Moro-Martin, A. \& Malhotra, R. 2002, \aj, 124, 2305
\bibitem[Ozernoy \etal(2000)]{oz00} Ozernoy, L. M., Gorkavyi, N. N., Mather, J. C., \& Taidakova, T. A. 2000, \apj, 537, L147
\bibitem[Quillen \& Thorndike(2002)]{qt02} Quillen, A. C., \& Thorndike S. 2002, \apj, 578, L149
\bibitem[Wilner \etal (2002)]{whk02} Wilner, D. J., Holman, M. J., Kuchner, M. J., \& Ho, P. T. P. 2002, \apj, 569, L115
\bibitem[Wisdom \& Holman(1991)]{wisdom91} Wisdom, J. \& Holman, M, 1991, \aj, 102, 1528
\bibitem[Wyatt \& Dent(2002)]{wyatt02} Wyatt, M. C. \& Dent. W. R. F. 2002, \mnras, 334, 589
\bibitem[Wyatt \& Whipple(1950)]{wyatt50} Wyatt, S. P. \& Whipple, F. L. 1950, \apj, 111, 134
\bibitem[Zuckerman(2001)]{zuc01} Zuckerman, B. 2001, \araa, 39, 549
\end{thebibliography}
\end{document}